\documentclass[a4paper,9pt]{article}

\usepackage{amsmath, amssymb, amsthm} 
\usepackage{graphicx} 
\usepackage{hyperref} 
\usepackage{cite} 
\usepackage{bm}
\usepackage{algorithm}
\usepackage{algpseudocode}
\usepackage{dsfont}
\usepackage{graphicx}
\usepackage{float}
\usepackage{subcaption}
\usepackage{geometry} 
\usepackage[numbers]{natbib}

\title{The Spatial Regime Conversion Method}
\author{
Charles G. Cameron, Cameron A. Smith,  Christian A. Yates \\ \small{Centre for Mathematical Biology, Department of Mathematical sciences, University of Bath}}

\begin{document}

\maketitle
\begin{abstract}
We present the spatial regime conversion method (SRCM), a novel hybrid modelling framework for simulating reaction-diffusion systems that adaptively combines stochastic discrete and deterministic continuum representations. Extending the regime conversion method (RCM) to spatial settings, the SRCM employs a discrete reaction-diffusion master equation (RDME) representation in regions of low concentration and continuum partial differential equations (PDEs) where concentrations are high, dynamically switching based on local thresholds. This enables efficient and accurate simulation of systems in which stochasticity plays a key role but is not required uniformly across the domain. We specify the full mathematical formulation of the SRCM, including conversion reactions, hybrid kinetic rules, and consistent numerical updates. The method is validated across several one-dimensional test systems, including simple diffusion from a region of high concentration, the formation of a morphogen gradient, and the propagation of FKPP travelling waves. Results show that the SRCM captures key stochastic features while offering substantial gains in computational efficiency over fully stochastic models.

\end{abstract}
\vspace{0.5em}
\hrule
\vspace{1em}

\section{Introduction}

Reaction-diffusion models form a fundamental class of mathematical frameworks widely used to describe biological and physical processes. By definition, these models capture both the diffusion of species and their kinetic interactions. Applications include wound healing \citep{callaghan_stochastic_2006}, where logistic growth kinetics are employed; the spread of infectious diseases, where classic compartmental infectious disease dynamics are extended spatially \citep{Volpert_Petrovskii_2009}; and the formation of spatial patterns in semi-arid ecosystems \citep{Klausmeier_1999,Sherratt_2013} where spatial inhomogeneities arise as a result of diffusion-driven instability \citep{Turing_1952}.

Depending on the required level of detail and spatial resolution, reaction-diffusion systems can be represented using several different mathematical approaches. At the microscopic scale, individual particles and their interactions are tracked explicitly in continuous space. Methods such as Brownian dynamics \citep{Andrews_Bray_2004} or Langevin dynamics \citep{langevin1908brownian} fall into this category. While these techniques are considered the most accurate representations, they can be computationally expensive — a feature typically attributed to the requirement of tracking exact pairwise interactions. Microscopic methods are particularly useful when the specific details of species interactions are important, for instance, in modelling calcium ion release from intracellular stores \citep{Dobramysl2016}, in which the specific timing and location of the direct binding of calcium ions determine the release of stored calcium.

At an intermediate, mesoscopic scale, compartment-based methods partition space into discrete compartments. Particles move between neighbouring compartments and react within them. These dynamics can be analysed through the reaction-diffusion master equation (RDME) and are typically simulated using stochastic simulation algorithms (SSAs). Gillespie's exact SSA \citep{Gillespie_1977} produces trajectories consistent with the RDME by assigning exponential waiting times to reaction and diffusion events. Approximate methods such as tau-leaping \citep{Gillespie_2001} reduce computational expense (at the cost of exactness) by executing multiple reactions within the same fixed time step. Compartmental stochastic techniques have been applied to biological systems including the p53-Mdm2 regulatory feedback loop \citep{Proctor_Gray_2008} and models of gene expression during cell growth and division \citep{Lu_Volfson_Tsimring_Hasty_2004}, offering a more efficient alternative to microscopic methods while still capturing important stochasticity.

At the macroscopic scale, continuous models typically describe species concentrations using partial differential equations (PDEs). These models effectively assume sufficiently high particle counts that we can ignore any stochastic fluctuations. However, this assumption is often violated, especially in biological systems for which stochastic effects can drive key behaviours such as switching and oscillation \citep{Noble_2021}. PDEs can be derived via moment closure approximations of the RDME \citep{Erban_Chapman_2019}, and while computationally efficient and suitable for some systems, they may fail to capture critical noise-driven dynamics. PDEs benefit from a wide range of established numerical schemes and offer analytical advantages, such as the ability to identify steady states and their stabilities, explore bifurcations, and perform tractable theoretical analysis.

Many biological systems are inherently multiscale; some parts of the domain have high copy-numbers while other areas have low copy-numbers. This means that some spatial regions require a more detailed model to correctly capture the dynamics, while in other areas, a less detailed model will suffice. This motivates the concept of spatially extended hybrid methods, which aim to combine the complementary advantages of coarse-grain and fine-grain methods whilst negating their complementary weaknesses. These frameworks may couple detailed models in regions where stochasticity is important with coarser models for regions in which fluctuations can be neglected. In particular, mesoscopic-macroscopic hybrid approaches offer a useful balance between computational efficiency and accuracy \citep{Smith_Yates_2018}. In regions with low particle numbers, stochastic compartment-based models are used; elsewhere, PDEs provide a deterministic description of concentration dynamics.

A typical use of such hybrid methods is to model travelling waves \citep{kolmogorov1937study,Moro_2004,Harrison_Yates_2016} in which the dynamics of the relatively low number of particles at the front determines the wave speed. However, behind the wave-front stochastic fluctuations are far less important to the overall dynamics and, as a result, it is sufficient to use a PDE for the average dynamics in that region. 
 
A key limitation of many existing mesoscopic-to-macroscopic hybrid methods is their reliance on a single (often fixed) interface between stochastic and deterministic regimes \citep{Yates_Flegg_2015}. These approaches typically lack versatility; require prior knowledge of where the interface should be placed; cannot easily handle multiple interfaces; and are poorly suited to irregular or non-planar geometries. 

In this paper, we introduce the spatial regime conversion method (SRCM), an adaptive hybrid framework that couples the mesoscale and macroscale, in order to combat these challenges. Our method dynamically converts mass between the two regimes and allows for mixed mass representation within each compartment. When concentrations are low (below a given threshold), the system converts continuous mass into discrete mass to be modelled stochastically. Conversely, when concentrations are high (exceeding a given threshold), mass is converted to a deterministic PDE representation. This adaptive conversion removes the need for user-defined interfaces and potentially makes the method particularly suitable for complex systems such as Turing pattern formation \citep{Turing_1952, Karig_Martini_Lu_DeLateur_Goldenfeld_Weiss_2018}, where interface location is \textit{a priori} unknown or varies over time. The SRCM offers computational efficiency by limiting stochastic simulation only to the regions in which it is truly required.

Our work builds on the regime conversion method (RCM) introduced by \citet{Kynaston_Yates_Hekkink_Guiver_2023}, which couples discrete and continuous descriptions of species reacting in a well-mixed system. Their framework uses conversion reactions to switch between representations based on species abundance. Thresholds and rate parameters control these conversions. An alternative approach, the jump-switch-flow (JSF) framework, instead treats each species as either fully discrete or fully continuous at any given time, with regime changes occurring when abundances cross a fixed threshold \citep{germano2024hybridframeworkcompartmentalmodels}.

In this paper, we extend the RCM \citep{Kynaston_Yates_Hekkink_Guiver_2023} to spatially distributed systems in one dimension. This generalisation requires careful treatment of local concentrations, regime transitions, and spatial coupling. Our method retains the adaptivity of the original RCM, enabling accurate and efficient simulation of systems for which stochasticity plays a fundamental role in the dynamics. By applying the discrete representation only where it is truly necessary, the method achieves a substantial gain in computational efficiency while maintaining accuracy in critical regions. This flexible framework is especially powerful for modelling systems with heterogeneous spatial features.

In section 2, we introduce the methodological background, describing how PDEs and compartment-based stochastic methods can be used to model reaction-diffusion systems. In section 3, we  establish the requisite notation and present the simulation algorithm for a purely stochastic canonical method first. We then describe how we augment the canonical stochastic system and allow it to interface with the continuous PDE in order to construct the full SRCM. In section 4, we apply the SRCM to four different reaction-diffusion test problems. We compare the resulting solution profiles to those obtained from purely deterministic and purely stochastic simulations, and assess the relative computational efficiency of the method.

Note, in what follows, straight letters will normally denote the name or label of a species or reaction, while italicised letters will usually denote a numerical quantity.

\section{Methodological background}

We consider the spatiotemporal evolution of a collection of chemical species labelled \(\mathbf{A} = [\text{A}^{(1)}, \dots, \text{A}^{(M)}]\), within a one-dimensional spatial domain \(\Omega = [0, L]\) and over a finite time interval \(t \in [0, T]\). The dynamics of the system are governed by a set of chemical reactions \(R\), which, together with the species \(\mathbf{A}\), define a chemical reaction network, \(\mathcal{N}\).

\subsection{Compartment-Based Modelling}

We first introduce a compartment-based modelling framework for the chemical reaction network. The domain \(\Omega\) is discretised into \(K \in \mathbb{N}\) non-overlapping compartments, denoted by \(\{\Omega_k\}_{k=1}^{K}\), such that $\Omega=\bigcup_{k=1}^K \Omega_k$. Each compartment has uniform length \(|\Omega_k|= L/K\) for simplicity, but uniformity is not strictly required.

For each species \(\text{A}^{(i)}\), we denote by \(N^{(i)}_k(t)\) the number of particles of that species present in compartment \(\Omega_k\) at time \(t\). The total number of particles of species \(\text{A}^{(i)}\) across the entire spatial domain at time \(t\) is then given by the following,
\[
N^{(i)}(t) := \sum_{k=1}^{K} N^{(i)}_k(t).
\]
We denote the state of the system at time \(t\) by a matrix \(\boldsymbol{N}(t) \in \mathbb{N}_0^{M \times K}\), where the matrix entry \(N_{i,k}(t) := N^{(i)}_k(t)\), represents the number of particles of species \(i \in \{1, \dots, M\}\) in compartment \(\Omega_k\), for each \(k \in \{1, \dots, K\}\). This formulation provides a compact representation for tracking the spatial distribution of all species across the entire domain.

We consider initially two types of reactions: kinetic reactions, denoted \(\text{R}^{\mathrm{kin}}\), and diffusion reactions, denoted \(\text{R}^{\mathrm{diff}}\). The kinetic reactions occur within each compartment, and the diffusion reactions allow particles to move between adjacent compartments. Each species in a given compartment is associated with two diffusion reactions: leftward and rightward movement. We define the set of diffusion reaction labels as
\[
\text{R}^{\mathrm{diff}} = \left\{ \text{d}^{(1)}_{-}, \dots, \text{d}^{(M)}_{-}, \text{d}^{(1)}_{+}, \dots, \text{d}^{(M)}_{+} \right\},
\]
where the superscript indicates the species undergoing diffusion, and the subscript \(-\) or \(+\) denotes the direction of movement (left or right, respectively). We assume that each species diffuses with a constant diffusivity throughout the domain, but which may vary by species. The first \(M\) diffusion reactions listed in $\text{R}^{\mathrm{diff}}$ correspond to movement to the left for each species, while the remaining \(M\) describe movement to the right.

In addition to diffusion, we consider \(Q\) kinetic reactions in each compartment. The corresponding set of kinetic reaction labels is given by:
\[
\text{R}^{\mathrm{kin}} = \{ \text{r}_1, \cdots, \text{r}_Q \}.
\]
In this paper we will work only with systems that are at most second-order, although higher-order reactions could be incorporated without much conceptual difficulty. The set of kinetic reactions \(\text{R}^{\mathrm{kin}}\) is correspondingly partitioned into three disjoint subsets corresponding to zeroth-, first-, and second-order reactions:
\[
\text{R}^{\mathrm{kin}} = \text{R}^{\mathrm{kin}}_0 \cup \text{R}^{\mathrm{kin}}_1 \cup \text{R}^{\mathrm{kin}}_2,
\]
where \(|\text{R}^{\mathrm{kin}}_0| = Q_0\), \(|\text{R}^{\mathrm{kin}}_1| = Q_1\), and \(|\text{R}^{\mathrm{kin}}_2| = Q_2\), so that the total number of kinetic reactions is \(Q = Q_0 + Q_1 + Q_2\). For each compartment \(\Omega_k\), where \(k \in \{1, \dots, K\}\), we define the local reaction set as follows:

\[
\text{R}_k = \{ \text{R}^{\mathrm{diff}} , \text{R}^{\mathrm{kin}}\},
\]
which comprises \(2M + Q\) total reactions: \(2M\) diffusion reactions (one left and one right for each species), and \(Q\) kinetic reactions.

Each reaction \(q \in \{1, \dots, 2M + Q\}\), when occurring in compartment \(\Omega_k\) (for \(k \in \{1, \dots, K\}\)) induces a change in the system state \(\boldsymbol{N}(t)\), which we encode with a stoichiometric update matrix \(\nu_{q,k} \in \mathbb{Z}^{M \times K}\). The collection \(\nu = (\nu_{q,k})_{q=1,\dots,2M+Q;\,k=1,\dots,K}\) defines a family of such update matrices. For any particular reaction $q^* \in \{1,\dots,2M+Q \}$ occurring in compartment $\Omega_{k^*}$ for $k^* \in \{1,\dots,K \}$, the \(\ell\)th column, \(\nu_{q^*,k^*}^{\ell} \in \mathbb{Z}^M\), specifies the change in copy numbers of all species in compartment \(\Omega_\ell\) resulting from reaction \(q^*\) occurring in compartment \(\Omega_{k^*}\). If the reaction is a kinetic reaction occurring in compartment $\Omega_{k^*}$, then $\nu_{q^*,k^*}^{\ell}=\bf{0}$ for $\ell\neq k^*$ since kinetic reactions only affect species numbers within their own compartment. For diffusion reactions from compartment $\Omega_{k^*}$, we have that $\nu_{q^*,k^*}^{\ell}=\bf{0}$ for $\ell\neq k^*,k^*\pm 1$, depending on the direction of movement. 


We denote by \(\alpha_{q,k}\) the propensity of reaction \(q\) occurring in compartment \(\Omega_{k}\). This propensity depends on the current state of the system, and is typically determined by the number of molecules within compartment $\Omega_{k}$ involved in reaction $q$.

Simulation of the system is performed using the Gillespie algorithm \citep{Gillespie_1977} in which the waiting time until the next reaction, $\tau$, is drawn from an exponential distribution with mean inversely proportional to the sum of the propensities across every reaction and compartment. The algorithm then selects both the compartment and the reaction to occur, based on these propensity values. If reaction \(q\) is chosen to occur in compartment \(\Omega_{k}\), the system state is updated by applying the corresponding stoichiometric change:
\[
\boldsymbol{N}(t+\tau) = \boldsymbol{N}(t) + \nu_{q,k}.
\]

\subsection{Partial differential equation based modelling}

We can model reaction-diffusion systems using PDEs, describing an approximation of the mean concentration of species derived from the moment closure of the RDME \citep{Erban_Chapman_2019}. Let \( u^{(i)}(x, t) \) denote the concentration of species \( i \in\{1,\dots,M\}\) at location \( x\in\Omega \) and time \( t\in[0, T] \). We define the general form for this PDE for species $u^{(i)}$ as follows:

\begin{equation}
\frac{\partial u^{(i)}}{\partial t} = D^{(i)}\frac{\partial^2 u^{(i)}}{\partial x^2} + f^{(i)}(\mathbf{u}),\label{eqution:PDE_species_i}
\end{equation}
where $f^{(i)}(\mathbf{u})$ encodes the reaction kinetics of the system and $D^{(i)}$ is the diffusion coefficient of species $i$.

Only under very specialised conditions can the solutions to such PDEs be obtained analytically. Consequently, we will resort to finding the numerical solution of such systems. A numerical approach will also make sense when we come to interfacing our PDE solutions with our RDME simulations in the SRCM.

To solve PDE \eqref{eqution:PDE_species_i} numerically, we discretise the domain into \( J+1 \) spatial grid points (inclusive of the two boundaries) with spacing \( \Delta x = L/J \). We choose \( J > K \), ensuring a finer spatial resolution of the PDE solution relative to the compartmental model. We set the number of grid points to be an integer multiple of the number of compartments, $J=PK$, where $P > 1$ is the resolution factor. We use $\Delta t$ to denote the temporal spacing. The spatial and temporal grid points are indexed as follows \citep{MortonMayers2005}:

\[
x_j = j\Delta x, \quad j= \{0,1,\cdots,J\}\quad \text{and}\quad t_n=n\Delta t, \ n \in \mathbb{N}_0.
\]
 The numerical solution $u^{(i)}_j(t_n)$ approximates the solution $u^{(i)}(x_j, t_n)$ of PDE \eqref{eqution:PDE_species_i}:
\[
u^{(i)}_j(t_n) \approx u^{(i)}(x_j, t_n).
\]
Note that in this setting, each \( x_j \) will be associated with a compartment \( \Omega_k \), providing a mapping between the PDE and compartment-based representations. For each $\Omega_k$ we use the left-hand-rule convention that; $x_j \in \Omega_k$ for $j \in \{(k-1)P,\cdots,kP-1\}$. In our implementation of the SRCM we will be using a finite-difference-based fourth-order Runge-Kutta method to solve the PDE, although other suitable numerical solution methods can be substituted.

\section{The Spatial Regime Conversion Method}

Now that we have outlined the basis of the two reaction-diffusion modelling paradigms that we will employ in the SRCM we can outline how they are coupled together. 
In section \ref{Section:canonical_method} we discuss the simulation of a canonical RDME, which allows us to introduce the notational framework required for the general simulation of the SRCM. Building on this, we outline the mechanics and the algorithm for the SRCM (which adaptively combines PDE and RDME frameworks) in section \ref{section:SRCM}.

\subsection{Canonical stochastic method}\label{Section:canonical_method}
In this section we consider a canonical stochastic system, introducing the notation and general framework associated with the SRCM without explicitly interfacing with the PDEs (this will be done in section \ref{section:SRCM}). This canonical system will incorporate diffusion, reaction kinetics and also, importantly, the forward and backward conversion reactions (although these will not take effect until we interface with the PDE in section \ref{section:SRCM}). In the regime conversion method \citep{Kynaston_Yates_Hekkink_Guiver_2023} `forward' conversion is the conversion from continuous to discrete mass while the `backward' conversion refers to the conversion from discrete to continuous. We will refer to these as `C-D conversion' and `D-C conversion' to mean continuous-to-discrete and discrete-to-continuous conversion, respectively. At this point, while introducing the canonical stochastic framework these conversion reactions are simply placeholders --- dummy reactions with no impact on the system --- however, it is important to introduce them here as these reactions later control transitions between discrete and continuous representations in the SRCM. We define the set of conversion reactions within each compartment as follows:

\[
\text{R}^{\text{con}} = \{\mathcal{C}_{\text{CD}}^{(1)}, \cdots, \mathcal{C}_\text{CD}^{(M)}, \mathcal{C}_\text{DC}^{(1)}, \cdots, \mathcal{C}_\text{DC}^{(M)}\}.
\]
The first $M$ reactions correspond to continuous-to-discrete conversion reactions for species $1,\dots,M$, respectively, whilst the second $M$ reactions correspond to discrete-to-continuous reactions for species $1,\dots,M$, respectively. 
We now extend the full reaction list for compartment $\Omega_k$ to include these conversion reactions:
\[
\text{R}_k = \{\text{R}^{\mathrm{con}},\text{R}^{\mathrm{diff}}, \text{R}^{\mathrm{kin}} \}.
\]
Each element of \(\text{R}_k\) is a label for a specific reaction initiated within compartment $\Omega_k$. As there are \(2M + 2M + Q\) distinct reaction types per compartment and \(K\) compartments, the system comprises a total of \(K(4M+Q)\) reactions. For \( q \in \{1,\dots, M\} \), the reaction is a C-D conversion of species \( i \equiv q \). For \( q \in \{M+1,\dots, 2M\} \), it is a D-C conversion of species \( i \equiv q \ (\text{mod}\, M) \). Reactions with \( q \in \{2M+1,\dots, 3M\} \) represent leftward diffusion, and those with \( q \in \{3M+1,\dots, 4M\} \) represent rightward diffusion, both of species \( i \equiv q \ (\text{mod}\, M) \). Finally, for \( q \in \{4M+1,\dots 4M+Q\} \), the reaction corresponds to one of the \( Q \) kinetic reactions. Whilst compartments share the same set of reactions, \( \text{R}_k \), the propensity functions of the diffusion reactions in the boundary compartments (i.e., \( k = 1 \) and \( k = K \)) are adjusted to implement the appropriate boundary conditions \citep{Erban2007rbc} (typically, in the test simulations of section \ref{section:results}, these will be zero-flux boundary conditions). The complete list of reactions across all compartments of the domain is given by \( \text{R} = \{ \text{R}_1, \text{R}_2, \dots, \text{R}_K \} \).

We introduce $\lambda_{q,k}$ for $q \in \{1, \cdots, 4M+Q \}$ and $k \in \{1,\cdots, K\}$, which define the rate of reaction $q$ occurring in compartment $k$. In general, these reaction rates will be consistent across compartments.

In the following subsections we will introduce notation for each of the three reaction types - conversion, diffusion, and kinetic, and discuss their implementation in the canonical stochastic framework.

\subsubsection{Conversion reactions}\label{conversion-reactions}

Here we briefly discuss conversion reactions, which handle transformations between discrete and continuous representations of species in the SRCM. Although conversion plays a crucial role in the full SRCM framework, it is important to stress that we have not yet discussed the inclusion of the PDE dynamics, so the aim of this subsection is to lay the groundwork for its later inclusion.

Let $q_1 \in \{1,\dots, M\}$ index a continuous-to-discrete conversion reaction of species \( i \), representing the transformation of a single particle's worth of continuous mass into a discrete particle. In the SRCM, this would describe mass transfer from the continuous to the discrete regime. For now, in this purely stochastic setting, we model this simply as a trivial first-order reaction (which does not impact the system):
\[
\text{A}^{(i)}_k \xrightarrow{\lambda_{q_1,k}} \text{A}^{(i)}_k.
\]
This results in an update to the discrete state in compartment \( k \):
\[
\nu_{q_1,k}^{\ell = k} = \mathbf{0} \in \mathbb{Z}^{M}, \quad \nu_{q_1,k}^{\ell \neq k} = \mathbf{0}\in \mathbb{Z}^{M}.
\]
Conversely, for the discrete-to-continuous conversion, suppose $q_2 \in \{M+1, \dots, 2M\}$. This would correspond to the conversion of a single discrete particle to particle's worth of continuous mass in the SRCM. However in the canonical stochastic system we simply represent this as:
\[
\text{A}^{(i)}_k \xrightarrow{\lambda_{q_2,k}} \text{A}^{(i)}_k,
\]
with the corresponding stoichiometric change:
\[
\nu_{q_2,k}^{\ell = k} = \mathbf{0} \in \mathbb{Z}^{M}, \quad \nu_{q_2,k}^{\ell \neq k} = \mathbf{0}\in \mathbb{Z}^{M}.
\]
Although these reactions currently have no effect on any discrete or continuous species (since such a representation is not yet included), they play a vital role in the SRCM.

\subsubsection{Diffusion reactions}\label{diffusion-reactions}

Let reaction \( q_3 \in \{2M+1,\dots,3M\} \) index the leftward diffusion of species \( \text{A}^{(i)} \). This reaction corresponds to the movement of a particle of species \( i \) from compartment \( k \) to compartment \( k - 1 \). While this reaction is defined for \( k \in \{1,\dots, K\} \) the corresponding propensity function for $k=1$ will be set to zero for the examples in this paper, implementing zero-flux boundary conditions by ensuring there is no leftward diffusion from the first compartment.

The stoichiometric update for this reaction is encoded in the matrix \( \nu_{q_3,k} \). Since only species \( i \) is involved, all rows except row \( i \) are zero. Similarly, since the diffusive jump affects only two compartments, the majority of the columns of \( \nu_{q_3,k} \) are zero. The non-zero columns are:
\[
\nu_{q_3,k}^{\ell = k} = -\bm{e}_i \in \mathbb{Z}^{M}, \quad \nu_{q_3,k}^{\ell = k - 1} = \bm{e}_i \in \mathbb{Z}^{M},
\]
where \( \bm{e}_i \in \mathbb{Z}^{M} \) is the standard basis vector with a 1 in the \( i \)-th position (and zeros elsewhere). For all other compartments:

\[
\nu_{q_3,k}^{\ell \neq k,k+1}  = \mathbf{0} \in \mathbb{Z}^{M}.
\]
Similarly, \( q_4 \in \{3M+1,\dots,4M\} \) indexes the rightward diffusion reactions, where species \( i \) moves from compartment \( k \) to compartment \( k + 1 \). While this reaction is defined for \( k \in \{1,\dots, K\} \) the corresponding propensity function for $k=K$ will be set to zero for the examples in this paper, implementing zero-flux boundary conditions by ensuring there is no rightward diffusion from the last compartment.

The corresponding stoichiometric update vectors are:
\[
\nu_{q_4,k}^{\ell = k} = -\bm{e}_i \in \mathbb{Z}^{M}, \quad \nu_{q_4,k}^{\ell = k + 1} = \bm{e}_i \in \mathbb{Z}^{M},
\]
and for all other compartments:
\[
\nu_{q_4,k}^{\ell \neq k,k+1}  = \mathbf{0} \in \mathbb{Z}^{M}.
\]

\subsubsection{Kinetic reactions}\label{kinetic-reactions}

We now focus on the kinetic reactions, which are local to each compartment. To represent these reactions, we define a stoichiometric vector \( \mu_q \in \mathbb{Z}^M \) for each reaction \( q \), where the scalar \( \mu_q^{(m)} \) denotes the net change in the number of particles of species \( m \) given reaction \( q \) occurs. Only the entries of \( \mu_q \) corresponding to the species involved in the reaction (as reactants or products) will be non-zero\footnote{Even some of these entries may be zero if the corresponding reactant acts as a catalyst and therefore emerges from the reaction unchanged in copy number.}.

Suppose a kinetic reaction with index \( q \in \{4M+1,\dots, 4M+Q\} \) occurs in compartment \( k \in \{1,\dots, K \} \). The vector \( \mu_q \) comprises the \( k \)-th column of the stoichiometric matrix $\nu_{q,k}$, while all other columns are zero:
\[
\nu_{q,k}^{\ell = k} = \mu_q \in \mathbb{Z}^{M};
\]

\[
\nu_{q,k}^{\ell \neq k} = \mathbf{0} \in \mathbb{Z}^{M}.
\]
Later, when we extend this notation to the full SRCM framework, it will be necessary to introduce new reaction types that allow interactions between discrete particles and continuous mass.

\subsubsection{Propensity functions}\label{propensity-functions}

We define \(\alpha_{q,k}\) to be the propensity function of reaction \(q\) occurring within compartment \(k\). We define 
\[
\bm{\alpha}_k = [\alpha_{1,k}, \cdots, \alpha_{4M+Q,k}]^T,
\]
to be the vector of all propensity functions occurring in compartment \(k\), for $k\in\{1,\dots, K\}$. We then concatenate these vectors in order to define the full propensity matrix across all compartments as follows:
\[
\bm{\alpha} = [\bm{\alpha}_1, \cdots, \bm{\alpha}_K].
\]
These propensity values are used in the Gillespie algorithm to determine both the timing and location of the next reaction event. We define the total propensity within each compartment:
\[
\sigma_k = \sum_{q = 1}^{4M+Q} \alpha_{q,k},
\]
and collect these totals into a vector:
\[
\bm{\sigma} = [\sigma_1, \cdots, \sigma_K],
\]
which will be used to determine the compartment index. We then calculate the total propensity sum that will be used to calculate the time for the next reaction:
\[
\Sigma = \sum_{i=1}^{K} \sigma_i.
\]
The Gillespie algorithm for this canonical stochastic reaction network is given in algorithm 1.

\begin{algorithm}[H]
\caption{Stochastic simulation of canonical system}
\begin{algorithmic}[1]

\State Set $t \gets 0$ \Comment{Initialise time}
\State Set $T$ \Comment{Set end time}
\State Initialise state vector $\bm{N}(0) \in \mathbb{N}^{M \times K}$

\While{$t \leq T$}
    \State Draw random numbers $r_1, r_2, r_3 \sim \text{uniform}(0,1)$

    \For{$k = 1$ to $K$} \Comment{Work out propensity list for every compartment}
        \State Compute propensities $\bm{\alpha}_k = [\alpha_{1,k}, \ldots, \alpha_{Q+4M,k}]$
        \State Compute total propensity in compartment $k$: $\sigma_k = \sum_{q=1}^{Q+4M} \alpha_{q,k}$
    \EndFor

    \State Compute total system propensity: $\Sigma = \sum_{i=1}^{K} \sigma_i$
    \State Compute time to next reaction: $\tau = \frac{1}{\Sigma} \ln\left(\frac{1}{r_1}\right)$

    \State Determine the reaction compartment, $k$, such that:
    \[
    \frac{1}{\Sigma} \sum_{i=1}^{k-1} \sigma_i \leq r_2 < \frac{1}{\Sigma} \sum_{j=1}^{k} \sigma_j
    \]

    \State Determine the reaction index, $q$, that occurs within compartment $k$ such that:
    \[
    \frac{1}{\sigma_k} \sum_{j=1}^{q-1} \alpha_{j,k} \leq r_3 < \frac{1}{\sigma_k} \sum_{j=1}^{q} \alpha_{j,k}
    \]

    \State Execute reaction $q$ in compartment $k$ by updating the state: $\bm{N}(t + \tau) = \bm{N}(t) + \nu_{q,k}$
    \State Advance time: $t \gets t + \tau$

\EndWhile

\end{algorithmic}
\end{algorithm}

\newpage
\subsection{Including the continuous representation - the Spatial Regime Conversion Method}\label{section:SRCM}

After developing the groundwork for the canonical stochastic model, we now interface this with the continuous representation to characterise the SRCM. To capture discrete and continuous dynamics, we extend the state space to include both representations for each species \( \text{A}^{(i)} \).

 We define the function $v^{(i)}(x,t)$ to be the continuous profile of species $i$ within the SRCM framework. In order to allow the interaction and conversion between continuous and discrete mass we will need to calculate the amount of continuum mass within each compartment and to be able to treat it, in some sense, as if it were discrete mass (i.e. particles). To that end (as in \citep{Yates_Flegg_2015}) we define the associated `pseudo-particle' count for species $i$ in compartment \( \Omega_k \) as \( Y^{(i)}_k \in \mathbb{R}_+ \) for $i\in\{1,\dots,M\}$ and $k\in\{1,\dots,K\}$.  The number of continuous-mass pseudo-particles in compartment \( \Omega_k \) is calculated as follows
\[
Y^{(i)}_k(t) := \int_{\Omega_k} v^{(i)}(x,t) \, dx.
\]
Since these values represent continuous mass, they may take non-integer values. The continuous representation of species \( \text{A}^{(i)} \) will be labelled \( \text{C}^{(i)}  \).

Correspondingly, the discrete representation of species \( \text{A}^{(i)} \) is labelled \( \text{D}^{(i)} \), with the number of discrete particles in compartment \( \Omega_k \) given by \( X^{(i)}_k \in \mathbb{N}_0 \). These values represent actual particle counts and are therefore restricted to non-negative integers.

We calculate the total number of particles of species \( \text{A}^{(i)} \) in compartment \( \Omega_k \) as the sum of its continuous and discrete components:
\[
N^{(i)}_k := X^{(i)}_k + Y^{(i)}_k.
\]

For the SRCM we must extend the chemical reaction network \( \mathcal{N} \) to incorporate both continuous and discrete species and the potential reactions between them. We label the extended chemical reaction network as \( \bar{\mathcal{N}} \). This extended network explicitly includes reactions involving both continuous and discrete components. Consequently, the theoretical framework and associated stoichiometric matrices must be adapted to account for transformations involving continuously represented mass.

In the SRCM, we extend the original set of reactions to a new, augmented reaction set:
\[
\bar{\text{R}}_k = [\text{R}^{\text{con}}, \text{R}^{\text{diff}}, \bar{\text{R}}^{\text{kin}}_{\text{D}}],
\]
where $\bar{\text{R}}^{\text{kin}}_{\text{D}}$ denotes the extended kinetic reaction set in the  discrete regime of the SRCM (which we will describe in section \ref{Kinetic reaction}). The conversion and diffusion reactions remain unchanged. Recall that, while the conversion reactions were redundant in the canonical stochastic model, they play a crucial role in the full SRCM, facilitating mass transfer between regimes in a concentration-dependent manner (as described in section \ref{subsection:conversion_reactions}).

To describe the system state in this extended framework, we introduce the new state matrix $\bar{\boldsymbol{N}} \in \mathbb{N}^{2M \times K}$. The first $M$ rows of the state matrix represent the number of discrete particles, and the second $M$ rows represent the number of continuous pseudo-particles. Formally, we define:
\[
\bar{N}_{i,k} =
\begin{cases}
X^{(i)}_k, & \text{if } i \in \{1,\dots, M\}, \\
Y^{(i - M)}_k, & \text{if } i \in \{M+1,\dots, 2M\},
\end{cases}
\quad \forall\ k \in \{1,\dots,K \}.
\]

We now turn to defining the stoichiometric updates in the extended system. In the canonical system, each reaction was associated with a stoichiometric matrix $\nu_{q,k} \in \mathbb{Z}^{M\times K}$. In the SRCM, we must extend this to a stoichiometric matrix $\bar{\nu}_{q,k} \in \mathbb{Z}^{2M\times K}$, capturing updates to both the discrete and continuous components of the system state.

\subsubsection{Updating the continuous profile}

As described previously, in the event that a reaction occurs which changes the number of continuous pseudo-particles, with the corresponding state update equation:
\[
\bar{\bm{N}}(t+\tau) = \bar{\bm{N}}(t)+\bar{\nu}_{q,k},
\]
there will be some non-zero element within the final $M$ rows of $\bar{\nu}_{q,k}$. This will represent an integer change $\xi$ in at least one of the species $i \in \{1,\dots,M\}$ which is captured by the following equation for the number of pseudo-particles of species $i$ in compartment $\Omega_k$:
\begin{equation}\label{Cont change}
  Y^{(i)}_k(t+\tau) = Y^{(i)}_k(t)+\xi.
\end{equation}
We need to update the continuous representation of the species by adding a (possibly negative) mass $\xi$, spread across the area that corresponds to compartment $\Omega_k$. This change would be implemented as follows, in the continuous PDE solution:

\begin{equation}
v^{(i)}(x, t+\tau) = v^{(i)}(x, t) + \frac{\xi}{|\Omega_k|} \cdot \mathbf{1}_{\Omega_k}(x),
\end{equation}
where \( \mathbf{1}_{\Omega_k}(x) \) is the indicator function for compartment \( \Omega_k \), and \( |\Omega_k| \) denotes its volume.

However, since we are implementing a numerical solution of the PDE, we add $\xi$ particles' worth of mass spread evenly across the grid points in the associated compartment (remembering we are mapping between PDE points and compartments using the left-hand rule):

\begin{equation}
v_{j}^{(i)}(t+\tau) = v_j^{(i)}(t) + \frac{\xi}{|\Omega_k|},  \ \forall j \in\{(k-1)P,\cdots,kP-1\}.
\end{equation}
In the following subsections, we extend the conversion, diffusion and kinetic reactions for the SRCM in order to demonstrate how the framework changes with respect to the canonical stochastic system outlined in section \ref{Section:canonical_method}.

\subsubsection{Conversion reactions}\label{subsection:conversion_reactions}

In this section we will outline the implementation of conversion reactions in the full SRCM. Specifically we will consider a conversion reaction assumed to be occurring in compartment $\Omega_k$ for $k \in \{1,\dots,K \}$. In the canonical system, conversion reactions are described by the stoichiometric matrices \( \nu_{q_1,k} \) for C-D conversion reactions, where \( q_1 \in \{1, \dots, M\} \), and by \( \nu_{q_2,k} \) for D-C conversion reactions, where \( q_2 \in \{M+1, \dots, 2M\}\). Because we did not include the continuous representation in the purely stochastic canonical system, these reactions were essentially redundant.

In the SRCM, we define the augmented stoichiometric matrices \( \bar{\nu}_{q_1,k} \in \mathbb{Z}^{2M \times K} \), for the C-D conversion reactions. The first \( M \) rows correspond to the changes in the discrete representation, and the second \( M \) rows represent changes in their continuous counterparts. Specifically, the stoichiometric matrix for the C-D conversion reactions in compartment $\Omega_k$ are defined as follows:
\[
\bar{\nu}_{q_1,k}^{\ell = k} = \bm{e}_i - \bm{e}_{i+M} \in \mathbb{Z}^{2M}, \quad \bar{\nu}_{q_1,k}^{\ell \neq k} = \mathbf{0} \in \mathbb{Z}^{2M}.
\]
This results in the appropriate increase in the discrete particle numbers whilst also specifying that we should update the numerical PDE concentration profile as follows:
\begin{equation}
v_{j}^{(i)}(t+\tau) =v_j^{(i)}(t) - \frac{1}{|\Omega_k|},  \ \forall\quad j \in\{(k-1)P.\cdots,kP-1\}.
\end{equation}
Note that if $v^{(i)}_j(t) \leq \frac{1}{|\Omega_k|}$ for any PDE solution point, $j$, in the associated compartment, then this conversion will be aborted, as this would lead to negative values in the numerical PDE solution profile.

The stoichiometric matrix for the D-C conversion reaction, $q_2$ in compartment $\Omega_k$ is defined as follows:
\[
\bar{\nu}_{q_2,k}^{\ell = k} = -\bm{e}_i + \bm{e}_{i+M} \in \mathbb{Z}^{2M}, \quad \bar{\nu}_{q_2,k}^{\ell \neq k} = \mathbf{0} \in \mathbb{Z}^{2M}, \quad \text{for}\quad \ q_2 \in \{M+1,\dots,2M \}.\]
This results in the appropriate decrease in the discrete particle numbers as well as specifying that we should update the numerical PDE profile as follows:
\begin{equation}
v_{j}^{(i)}(t+\tau) = v_j^{(i)}(t) + \frac{1}{|\Omega_k|},  \ \forall\quad j \in\{(k-1)P,\cdots,kP-1\}.
\end{equation}
Note that there are no restrictions on D-C conversions since discrete particle numbers are integer valued; if the  number of particles of species $i$ is zero, that species will never be selected for a D-C conversion.

The conversion reactions are regulated by concentration thresholds, $\Theta_i$, for each species $i \in \{1, \dots, M\}$. These thresholds correspond to mass thresholds within a compartment, defined as $\Theta_i \cdot |\Omega_k|$, where $|\Omega_k|$ is the compartment volume. In a given compartment $\Omega_k$, if the mass of species $i$, denoted $N_k^{(i)}$, exceeds its mass threshold $\Theta_i \cdot |\Omega_k|$, then D-to-C conversion is active; otherwise, C-to-D conversion takes place.

We denote the full set of conversion reactions as  
\[
\text{R}^{\text{con}} = \{\mathcal{C}_{CD}^{(1)}, \dots, \mathcal{C}_{CD}^{(M)}, \mathcal{C}_{DC}^{(1)}, \dots, \mathcal{C}_{DC}^{(M)}\},
\]
which are included in the local reaction list $\text{R}_k$ for compartment $\Omega_k$. As before, we index these reactions by $q \in \{1, \dots, 2M\}$. The rate of each conversion reaction in compartment $k$ is defined as:

\[
\lambda_{q,k} = 
\begin{cases}
\gamma \cdot \mathds{1}_{\{N^{(q)}_k \leq \Theta_q \cdot |\Omega_k|\}}, & \text{for } q \in \{1, \dots, M\}, \\
\gamma \cdot \mathds{1}_{\{N^{(q-M)}_k >  \Theta_{q-M} \cdot |\Omega_k|\}}, & \text{for } q \in \{M+1, \dots, 2M\}.
\end{cases}
\]

\subsubsection{Diffusion reactions}

In the canonical system, the diffusion reactions are described by \( \nu_{q_3,k} \in \mathbb{N}^{M \times K} \) for leftward jumps, where \( q_3 \in \{2M+1, \dots, 3M\} \), and by \( \nu_{q_4,k} \in \mathbb{N}^{M \times K} \) for rightward jumps, where \( q_4 \in \{3M+1, \dots, 4M \} \). In the extended system, we define \( \bar{\nu}_{q_3,k} \in \mathbb{Z}^{2M \times K} \), where the first \( M \) rows are identical to \( \nu_{q_3,k} \) and the remaining \( M \) rows are zero. This is because diffusion reactions only impact the discrete regime and do not affect the continuous regime. The same extended structure applies to \( \bar{\nu}_{q_4,k} \).

\subsubsection{Kinetic reactions}\label{Kinetic reaction}

Kinetic reactions present potentially the greatest departure of the SRCM from the canonical stochastic system. They must be categorised by reaction order to be implemented appropriately in the SRCM. We classify reactions as zeroth-order, first-order and second-order. The assignment of products to either the discrete or continuous regime is governed by the following rules:

\begin{itemize}
  \item[(C1):] To minimise unnecessary transitions between modelling regimes, product molecules of species $\text{A}^{(i)}$ are assigned to the same regime as their corresponding reactants of species $\text{A}^{(i)}$ (if any). In the event of a homodimerisation reaction between $\text{D}^{(i)}$ and $\text{C}^{(i)}$ (discrete and continuous representation), we will place product molecules of this species in the discrete regime. 
  \item[(C2):] To preserve stochasticity, any reaction products that do not correspond to a reactant species are placed in the discrete regime.
  \item[(C3):] If all reactants in a reaction belong to the continuous regime, their products remain in the continuous regime.
  In this case, the reaction is naturally modelled using the PDE.
\end{itemize}
These rules were developed initially for the regime conversion method for well-mixed systems \citep{Kynaston_Yates_Hekkink_Guiver_2023}. In the spatial setting of the SRCM, we will adhere to these rules in the implementation of our test problems unless otherwise specified. 

In what follows, we consider kinetic reactions occurring within compartment $\Omega_k$, with the assumption that identical reactions are defined independently in other compartments. 

Given the stoichiometric vector $\mu_q \in \mathbb{Z}^M$ capturing the effect of reaction $q$ on the species numbers in compartment $\Omega_k$ in the canonical stochastic system, we define $\bar{\mu}_q \in \mathbb{Z}^{2M}$ to be the extended stoichiometric vector associated with reaction $q$ in which the first $M$ entries are the same as $\mu_q$ and the second set of $M$ entries are zero. We specify these as follows:

\begin{equation}
\bar{\mu}_q = \sum_{m=1}^M \bm{e}_m\mu_q^{(m)} \in \mathbb{Z}^{2M},\label{equation:mu_q}
\end{equation}
where $\bm{e}_m\in \mathbb{Z}^{2M}$ are $2M$-dimensional basis vectors and $\mu_q^{(m)}$ represents the $m^{\text{th}}$ element of $\mu_q$.

\subsubsection{Zeroth-order reactions}

Each zeroth-order reaction in the original discrete reaction system maps directly to a single corresponding reaction in the hybrid system.
A general zeroth-order reaction takes the form
\[
\emptyset \xrightarrow{\lambda_{q,k}} \text{A}^{(i)}_k,
\]
which will be implemented in the SRCM as follows:
\begin{align}
\emptyset &\xrightarrow{\lambda_{q,k}} \text{D}^{(i)}_k.\label{equation:zeroth_order}\tag{Z1}
\end{align}
In accordance with rule (C2), we assign all products to the discrete regime. The stoichiometric vector is therefore $\bar{\mu}_q$ as defined in equation \eqref{equation:mu_q}. Correspondingly, the stoichiometric matrix, $\bar{\nu}_{q,k}$, will have the first $M$ rows exactly the same as $\nu_{q,k}$ in the canonical stochastic system, and the second set of $M$ rows will be zero.

\subsubsection{First-order reactions}

In the canonical stochastic implementation we can write a generic first-order reaction as follows:
\[
\text{A}^{(i)}_k \xrightarrow{\lambda_{q,k}} (\mu_q^{(i)}+1)\text{A}^{(i)}_k + \sum_{\substack{m = 1\\ m \neq i}}^M \mu_q^{(m)} \text{A}^{(m)}_k,
\]
where $\mu_q \in \mathbb{Z}^M$ is the stoichiometric vector for the canonical system associated with reaction $q$. This corresponds to two possible implementations in the SRCM:
  \begin{align}
    \text{D}^{(i)}_k &\xrightarrow{\lambda_{q,k}} (\mu_q^{(i)}+1)\text{D}^{(i)}_k + \sum_{\substack{m=1\\m \neq i}}^M \mu_q^{(m)} \text{D}^{(m)}_k,\tag{F1}\label{equation:first_order_disrete} \\
    \text{C}^{(i)}_k &\xrightarrow{\lambda_{q,k}} (\mu_q^{(i)}+1)\text{C}^{(i)}_k + \sum_{\substack{m=1\\m \neq i}}^M \mu_q^{(m)} \text{C}^{(m)}_k.\label{equation:first_order_continuous}\tag{F2}
  \end{align}
Reaction \eqref{equation:first_order_continuous} involves only continuous species and is naturally implemented in the PDE regime. Consequently, we explicitly model only reaction \eqref{equation:first_order_disrete} within the discrete framework. Similarly to zeroth-order reactions, there is a one-to-one mapping between first-order reactions in the canonincal stochastic reaction system and reactions of type \eqref{equation:first_order_disrete} in the discrete regime of the SRCM. The stoichiometric vector is hence $\bar{\mu}_q$ as specified in equation \eqref{equation:mu_q}.

\subsubsection{Second-Order reactions}

Second-order reactions involve the interaction of two species and are written in general form as:
\[
\text{A}^{(i)}_k + \text{A}^{(j)}_k \xrightarrow{\lambda_{q,k}} (\mu_q^{(i)} + 1)\text{A}^{(i)}_k + (\mu_q^{(j)} + 1)\text{A}^{(j)}_k + \sum_{\substack{m = 1\\ m \neq i,j}}^M \mu_q^{(m)} \text{A}^{(m)}_k,
\]
where it is possible, in the case of homodimerisation, that $i=j$. This potentially gives rise to four distinct second-order reactions within the SRCM:
  \begin{align}
    \text{D}^{(i)}_k + \text{D}^{(j)}_k &\xrightarrow{\lambda_{q,k}} (\mu_q^{(i)}+1)\text{D}^{(i)}_k + (\mu_q^{(j)}+1)\text{D}^{(j)}_k + \sum_{\substack{m=1\\ m \neq i,j}}^M \mu_q^{(m)} \text{D}^{(m)}_k,\label{equation:S1} \tag{S1} \\
    \text{D}^{(i)}_k + \text{C}^{(j)}_k &\xrightarrow{\lambda_{q,k}} (\mu_q^{(i)}+1)\text{D}^{(i)}_k + (\mu_q^{(j)}+1)\text{C}^{(j)}_k + \sum_{\substack{m=1\\ m \neq i,j}}^M \mu_q^{(m)} \text{D}^{(m)}_k,\label{equation:S2} \tag{S2} \\
    \text{C}^{(i)}_k + \text{D}^{(j)}_k &\xrightarrow{\lambda_{q,k}} (\mu_q^{(i)}+1)\text{C}^{(i)}_k + (\mu_q^{(j)}+1)\text{D}^{(j)}_k + \sum_{\substack{m=1\\ m \neq i,j}}^M \mu_q^{(m)} \text{D}^{(m)}_k, \label{equation:S3} \tag{S3}\\
    \text{C}^{(i)}_k + \text{C}^{(j)}_k &\xrightarrow{\lambda_{q,k}} (\mu_q^{(i)}+1)\text{C}^{(i)}_k + (\mu_q^{(j)}+1)\text{C}^{(j)}_k + \sum_{\substack{m=1\\ m \neq i,j}}^M \mu_q^{(m)} \text{C}^{(m)}_k.
    \label{equation:S4} \tag{S4}
  \end{align}
As with the first-order reactions, the second-order interaction with purely continuous reactants \eqref{equation:S4} is implemented naturally in the PDE, whilst \eqref{equation:S1}, \eqref{equation:S2}, and \eqref{equation:S3} will be implemented in the discrete framework of the SRCM, in accordance with rule (C1). Second-order reactions yield greater complexity in the SRCM than zeroth- and first-order reactions.

In the canonical stochastic system, the total number of kinetic reactions is \( Q = Q_0 + Q_1 + Q_2 \), where the subscripts denote reaction order. In the extended hybrid system, each first-order reaction yields two subreactions (corresponding to whether the reactant is discrete or continuous), and each second-order reaction yields four (depending on the combination of discrete and continuous reactants involved), so the total number of kinetic reactions becomes:
\[
\bar{Q} = Q_0 + 2Q_1 + 4Q_2.
\]
Among these kinetic reactions, those with all reactants in the continuous state (\( \bar{Q}_C = Q_1 + Q_2 \)) are modelled in the continuous regime. The remaining:
\[
\bar{Q}_D = Q_0 + Q_1 + 3Q_2,
\]
are handled stochastically in the discrete regime. Let \( \bar{\text{R}}^{\text{kin}} \) denote the full set of kinetic reactions in the extended system, with:
\[
|\bar{\text{R}}^{\text{kin}}| = \bar{Q} = Q_0 + 2Q_1 + 4Q_2.
\]
We partition this set as:
\[
\bar{\text{R}}^{\text{kin}} = \bar{\text{R}}^{\text{kin}}_D \cup \bar{\text{R}}^{\text{kin}}_C,
\]
where \( \bar{\text{R}}^{\text{kin}}_D \) comprises the $\bar{Q}_D$ reactions with at least one discrete reactant, and \(\bar{\text{R}}^{\text{kin}}_C \) comprises the $\bar{Q}_C$ reactions with only continuous reactant species.

Diffusion and conversion reactions remain unchanged. Consequently, the complete set of reactions modelled in compartment $k$ of the discrete framework is:
\[
\bar{\text{R}}_k = [ \text{R}^{\text{con}}, \text{R}^{\text{diff}}, \bar{\text{R}}_D^{\text{kin}} ].
\]
We order the kinetic reaction set \( \bar{\text{R}}_D^{\text{kin}}\) in increasing reaction order: zeroth, first, then second, remembering that each second-order reaction of the canonical stochastic system yields three subreactions: \eqref{equation:S1}-\eqref{equation:S3}. We specify cumulative index sets in the extended reaction network. The zeroth-order reactions correspond to indices \( q \in \{4M + 1, \dots, 4M + Q_0\} \). The first-order reactions to indices \( q \in \{4M + Q_0 + 1, \dots, 4M + Q_0 + Q_1\} \). Finally, the second-order reactions correspond to indices \( q \in \{4M + Q_0 + Q_1 + 1, \dots, 4M + Q_0 + Q_1 + 3Q_2\} \), where the factor of three accounts for the three discrete sub-reactions associated with each second-order reaction.

We re-index kinetic reactions in the extended system using index \( q^* \). For zeroth- and first-order reactions, the mapping is identical:
\[
    q^* = q, \  \text{for} \ q \in \{4M+1, \dots, 4M + Q_0 + Q_1\}.
\]
Second-order reactions, originally indexed by
\[
q \in \{4M + Q_0 + Q_1 + 1, \dots, 4M + Q_0 + Q_1 + Q_2\},
\]
in the canonical stochastic system, are each expanded into three discrete sub-reactions,  \eqref{equation:S1}-\eqref{equation:S3}, in the extended system. Let
\[
i := q - (4M + Q_0 + Q_1),
\]
then the re-indexed sub-reactions are:
\[
q^*_L := 4M + Q_0 + Q_1 + 3(i - 1) + L, \quad \text{for} \quad L = 1,2,3 \quad \text{and} \quad \ i=1,\dots,Q.
\]

Given that $\text{R}^{\text{kin}} = \{\text{r}_1,\dots,\text{r}_{Q_0+Q_1+Q_2}\}$, then we will simply define $\bar{\text{R}}^{\text{kin}}_D = \{\bar{\text{r}}_1,\dots, \bar{\text{r}}_{Q_0+Q_1+3Q_2}\}$.

Consider a generic second-order reaction involving two species $\text{A}^{(i)}$ and $\text{A}^{(j)}$, within the SRCM. Each original second-order reaction \( q \) thus yields \( q^*_1, q^*_2, q^*_3 \), corresponding to reactions \eqref{equation:S1},\eqref{equation:S2} \eqref{equation:S3}, respectively. The associated stoichiometric vectors \( \bar{\mu}_{q^*_L} \in \mathbb{Z}^{2M} \) defined as:
\begin{align*}
\bar{\mu}_{q^*_1} &= \bar{\mu}_q := \sum_{m=1}^M \mu_q^{(m)} \bm{e}_m \in \mathbb{Z}^{2M},  \\
\bar{\mu}_{q^*_2} &= \mu_i \bm{e}_{i+M} + \sum_{\substack{m=1\\m \neq i}}^M \mu_q^{(m)} \bm{e}_m \in \mathbb{Z}^{2M}, \\
\bar{\mu}_{q^*_3} &= \mu_j \bm{e}_{j+M} + \sum_{\substack{m=1\\m \neq j}}^M \mu_q^{(m)} \bm{e}_m \in \mathbb{Z}^{2M}.
\end{align*}
Each kinetic reaction is implemented within a single compartment \( \Omega_k \), so the stoichiometric matrices are:
\[
\bar{\nu}_{q^*_L,k}^{\ell = k} = \bar{\mu}_{q^*_L}, \quad \bar{\nu}_{q^*_L,k}^{\ell \neq k} = \mathbf{0}.
\]
The full state update upon the firing of reaction $q_L^*$ in compartment $\Omega_k$ is given by:
\[
\bar{\bm{N}}(t+\tau) = \bar{\bm{N}}(t) + \bar{\nu}_{q^*_L,k}.
\]
If $L=2$ or $L=3$ then there will be an interaction between continuous and discrete regimes, and therefore we would like to update the amount of mass within the continuous PDE regime as follows:
\[
v^{(i)}(x, t+\tau) = v^{(i)}(x, t) + \frac{\bar{\mu}_{q^*_L}^{(i+M)}}{|\Omega_k|} \cdot \mathbf{1}_{\Omega_k}(x), \quad \text{for}\quad i = 1,\dots,M.\
\]
The vector $\bar{\mu}_{q^*_L}^{(i+M)}$ describes the change in the value $Y^{(i)}_k(t+\tau)$ given that reaction $\bar{r}_{q^*_L}$ has occurred in the extended system.
Correspondingly, we update the numerical solution of the PDE profile as follows:
\begin{equation}
v_{j}^{(i)}(t+\tau) = v_j^{(i)}(t) + \frac{\bar{\mu}_{q^*_L}^{(i+M)}}{|\Omega_k|},  \ \forall\  j \in\{(k-1)P,\cdots,kP-1\}.\nonumber
\end{equation}

\subsubsection{Propensity functions for extended system}\label{propensity-functions extended}

We define \(\bar{\alpha}_{q,k}\) as the propensity function of the reaction \(q\) that occurs within compartment \(\Omega_k\) of the SRCM. We note that the propensity functions will remain the same as the canonical system for $q \in \{1,\dots,4M+Q_0+Q_1\}$. However, we must introduce new propensity functions for $q \in \{4M+Q_0+Q_1+1,\dots,4M+Q_0+Q_1+3Q_2\}$, corresponding to the second-order interactions:
\[
\bar{\bm{\alpha}}_k = [ \alpha_{1,k}, \cdots, \alpha_{4M+Q_0+Q_1+3Q_2,k}]^T.
\]
The propensity functions of the second-order reactions may now depend on the continuous mass. For example, if we study reactions \eqref{equation:S1}--\eqref{equation:S3} then the associated propensities are as follows:
\begin{equation}
    \begin{aligned}
         \alpha_{q^*_1,k} = \lambda_{q^*_1,k}\cdot X^{(i)}_k \cdot X^{(j)}_k,\\
        \alpha_{q^*_2,k} = \lambda_{q^*_2,k}\cdot X^{(i)}_k \cdot Y^{(j)}_k, \\
        \alpha_{q^*_3,k} = \lambda_{q^*_3,k}\cdot  Y^{(i)}_k \cdot X^{(j)}_k. \\
    \end{aligned}
\end{equation}
Now that we have defined the full list of propensities for each compartment, we can concatenate them to produce the following propensity matrix:
\[
\bm{\bar{\alpha}} = [\bm{\bar{\alpha}_1}, \cdots, \bm{\bar{\alpha}_K}].
\]
We define the total propensity within each compartment as follows:
\[
\bar{\sigma}_k = \sum_{q = 1}^{4M+Q_0+Q_1+3Q_2} \bar{\alpha}_{q,k}.
\]
We collect these totals into a vector as follows:
\[
\bar{\sigma} = [\bar{\sigma}_1, \cdots, \bar{\sigma}_K],
\]
and define the overall propensity sum which will be used to calculate the time until the next reaction occurs:
\[
\bar{\Sigma} = \sum_{i=1}^{K} \bar{\sigma}_i.
\]

Throughout the simulations of the SRCM, we must update the state matrix $\bar{\bm{N}}(t) \in \mathbb{N}^{2M \times K}$ to reflect the number of pseudo-particles representing the continuous mass. This is essential because the propensity functions, specifically those associated with second-order mixed (discrete and continuous reactants) kinetic reactions and conversion reactions, depend on these values. In practice, however, we employ a more efficient strategy: rather than explicitly updating $\bar{\bm{N}}(t)$ at every PDE step, we compute the corresponding pseudo-particle numbers from the PDE solution whenever they are required by the stochastic dynamics. The pseudocode for the SRCM is presented in algorithm 2.

\begin{algorithm}[H]
  \caption{Spatial regime conversion method (SRCM)}
  \begin{algorithmic}[1]
    \State Set $t \gets 0$ \Comment{Initialise time}
    \State Set $T$ \Comment{Set end time}
    \State Initialise discrete state matrix ${\bm{N}}(0) \in \mathbb{N}^{M \times K}$
    \State Initialise PDE solution $v^{(i)}(x,0)$ for all $i \in \{1,\dots,M\}$
    \State Update extended state matrix $\bar{\bm{N}}(0)$ 
    \State $t_d = \Delta t$  \Comment{PDE timestep counter}

    \While{$t \leq T$}
      \For{$k = 1$ to $K$}
        \State Compute propensities $\bar{\alpha}_k = [\bar{\alpha}_{1,k}, \ldots, \bar{\alpha}_{4M+Q_0+Q_1+3Q_2,k}]$
        \State Compute total propensity for compartment $k$: $\bar{\sigma}_k = \sum_{q=1}^{4M+Q_0+Q_1+3Q_2} \bar{\alpha}_{q,k}$
      \EndFor

      \State Compute total system propensity: $\bar{\Sigma} = \sum_{k=1}^{K} \bar{\sigma}_k$
      \State Draw random numbers $r_1, r_2, r_3 \sim \text{uniform}(0,1)$
      \State Compute time to next reaction: $\tau = \frac{1}{\bar{\Sigma}} \ln\left(\frac{1}{r_1}\right)$

      \If{$t + \tau < t_d$} \Comment{Stochastic step}
        \State Determine compartment $k$ such that:
        \[
        \frac{1}{\bar{\Sigma}} \sum_{j=1}^{k-1} \bar{\sigma}_j \leq r_2 < \frac{1}{\bar{\Sigma}} \sum_{j=1}^{k} \bar{\sigma}_j
        \]

        \State Determine reaction $q$ within compartment $k$ such that:
        \[
        \frac{1}{\bar{\sigma}_k} \sum_{j=1}^{q-1} \bar{\alpha}_{j,k} \leq r_3 < \frac{1}{\bar{\sigma}_k} \sum_{j=1}^{q} \bar{\alpha}_{j,k}
        \]
        
        \If{$\bar{\nu}_{q,k}$ has non-zero continuous component}\Comment{A change in the continuous analogue}
          \For{$i = 1$ to $M$}
            \State $v^{(i)}_j(t+\tau) = v^{(i)}_j(t) + \frac{(\bar{\nu}_{q,k})_{(i,k)}}{|\Omega_k|}, \ \forall j \in \{((k-1)P,\dots,kP-1\}.$
          \EndFor
        \EndIf
        \State Update the extended state matrix $\bar{\bm{N}}(t+\tau)= \bar{\bm{N}}(t)+\bar{\nu}_{q,k}$.
        \State Advance time: $t \gets t + \tau$

      \Else \Comment{PDE step}
        \For{$i = 1$ to $M$}
          \State Perform an update step of the PDE for: $v^i_j(t + \Delta t)\ , \forall j \in \{0,\dots,PK\}$
        \EndFor
        \State Update the extended state matrix $\bar{\bm{N}}(t+\Delta t)$ following the PDE update.
        \State Set $t=t_d$
        \State Set $t_d = t+\Delta t$
      \EndIf
    \EndWhile
  \end{algorithmic}
\end{algorithm}

\section{Results}\label{section:results}

In this section we will consider four test problems designed to demonstrate the utility and versatility of the SRCM.
In order to characterise the accuracy of the SRCM we need to compare the resulting outcome of the combined mass in the SRCM against the equivalent quantity for the canonical stochastic simulations. 
Let \( \bm{N}_S(t) \in \mathbb{Z}^{M \times K} \) denote the state matrix of the canonical (fully stochastic) system at time \(t\) averaged over a number of repeats. 

Similarly, define $\bm{N}_H(t)$ to be the ensemble average of the SRCM state at time $t$ and $\bm{N}_P(t)$  to be the solution of the PDE at time $t$. As the PDE model is deterministic, it only needs to be run once, and thus no averaging is required. To enable direct comparison between the PDE and other models, we convert the PDE concentration profile into pseudo-particle counts over the compartmental domain. Throughout this section, We will define $h = |\Omega_k|$, to be the length of a compartment and $\mathcal{K} = \{1,\dots,K\}$ to be the indexing set of the compartments.

\subsection{Case 1: Diffusion from a top-hat initial condition} \label{Section:Test_case_1}

We will model a purely diffusive system using the SRCM for a single species, $\text{A}$, to first ensure that we can correctly simulate diffusion in the absence of kinetic reactions. As there is only one species we will drop the superscript denoting the species index for brevity. Correspondingly, $\textbf{A} = \{\text{A}\}$, and the reaction set is $\text{R}^{\mathrm{kin}} = \emptyset$, hence $\text{R}_k = [\text{R}^{\mathrm{con}},\text{R}^{\mathrm{diff}}] = [\mathcal{C}_{CD}, \mathcal{C}_{DC}, \text{d}_{-},\text{d}_{+} ]$. The corresponding mean-field PDE for this system is given by:
\begin{equation}
    \frac{\partial u}{\partial t} = D\frac{\partial ^2 u}{\partial x^2}, \quad x\in (0,L),\quad t\in(0,T].\label{equation:mean_field_diffusion_PDE}
\end{equation}
For simplicity, we choose to implement zero-flux boundary conditions:
\begin{equation}
\frac{\partial u}{\partial x}\Big|_{x=0,L}=0.\label{equation_zero_flux_BCs}
\end{equation}
The jump rate of the stochastic model, $d$, relates to the macroscopic diffusion coefficient $D$ as follows: $D = dh^2$. We use $K=50$ compartments and $\Omega = [0,1]$ so that $h=0.02$.
 We note that the solution of the PDE part of the SRCM, $v(x,t)$ evolves according to the same PDE and with the same boundary conditions as $u(x,t)$.

We run the SRCM and canonical stochastic models from a top-hat initial condition with all the mass initially represented in just two compartments in the centre of the domain in the stochastic regime. We run the PDE with the corresponding initial condition. The initial condition for the SRCM and the purely stochastic simulation is defined as:
\[
X_k(0) = 
\begin{cases}
100, \quad k\in \{24,25\}, \\
0, \quad k \in \mathcal{K}\setminus\{24,25 \},\\
\end{cases}\]
with
\[v(x,0) = \bm{0}
\] specified additionally for the initial condition of the PDE component of the SRCM.
The corresponding initial condition for the pure PDE is $u(x,0) = 5000 \cdot \mathbf{1}_{\Omega_{24} \cup\Omega_{25}}$, i.e. all mass is set within the central two compartments.

\begin{figure}[H]
    \centering
    \begin{subfigure}{0.4\textwidth}
        \includegraphics[width=\linewidth]{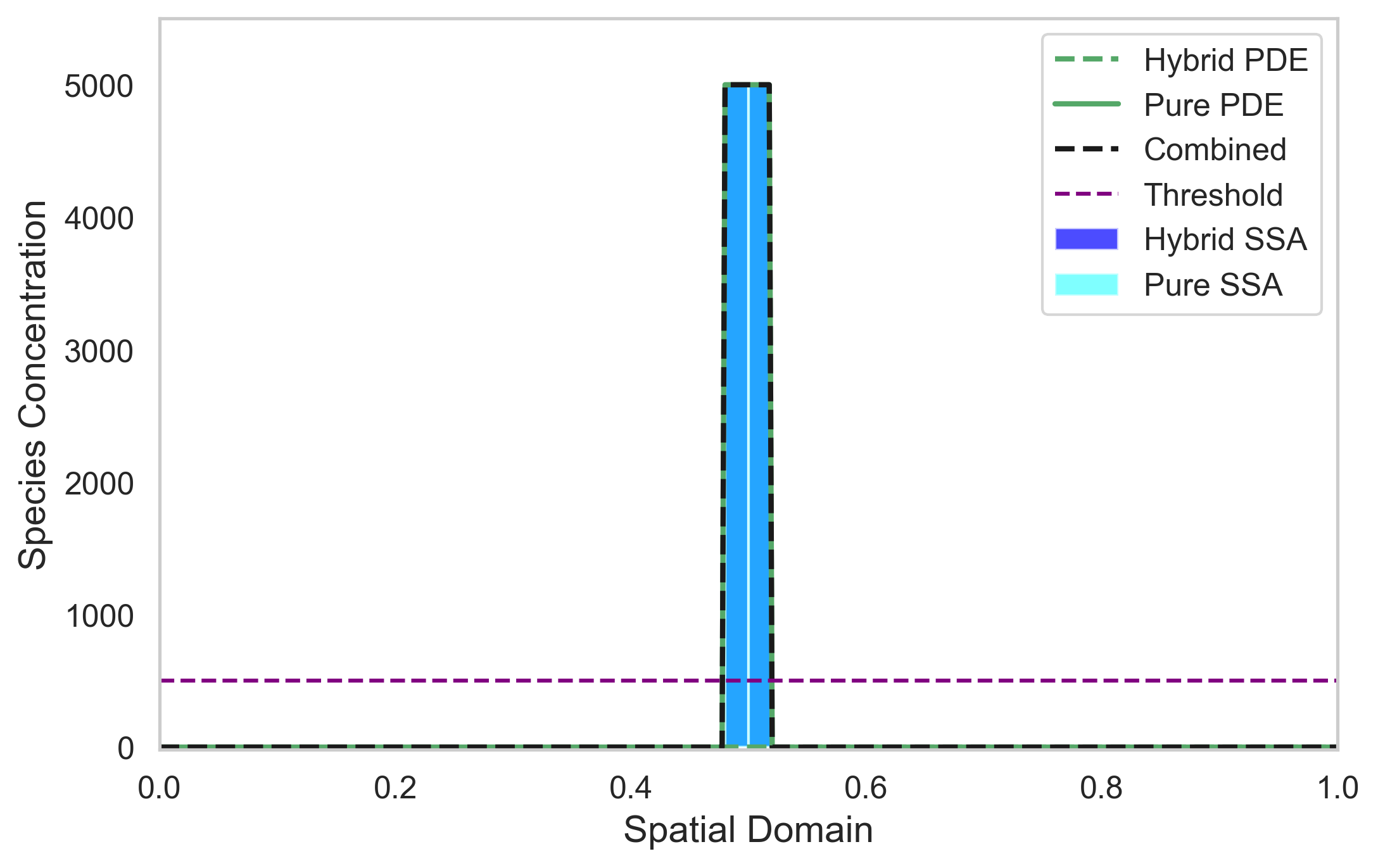}
        \caption{$t=0$}
        \label{fig:diff1}
    \end{subfigure}
    \begin{subfigure}{0.4\textwidth}
        \includegraphics[width=\linewidth]{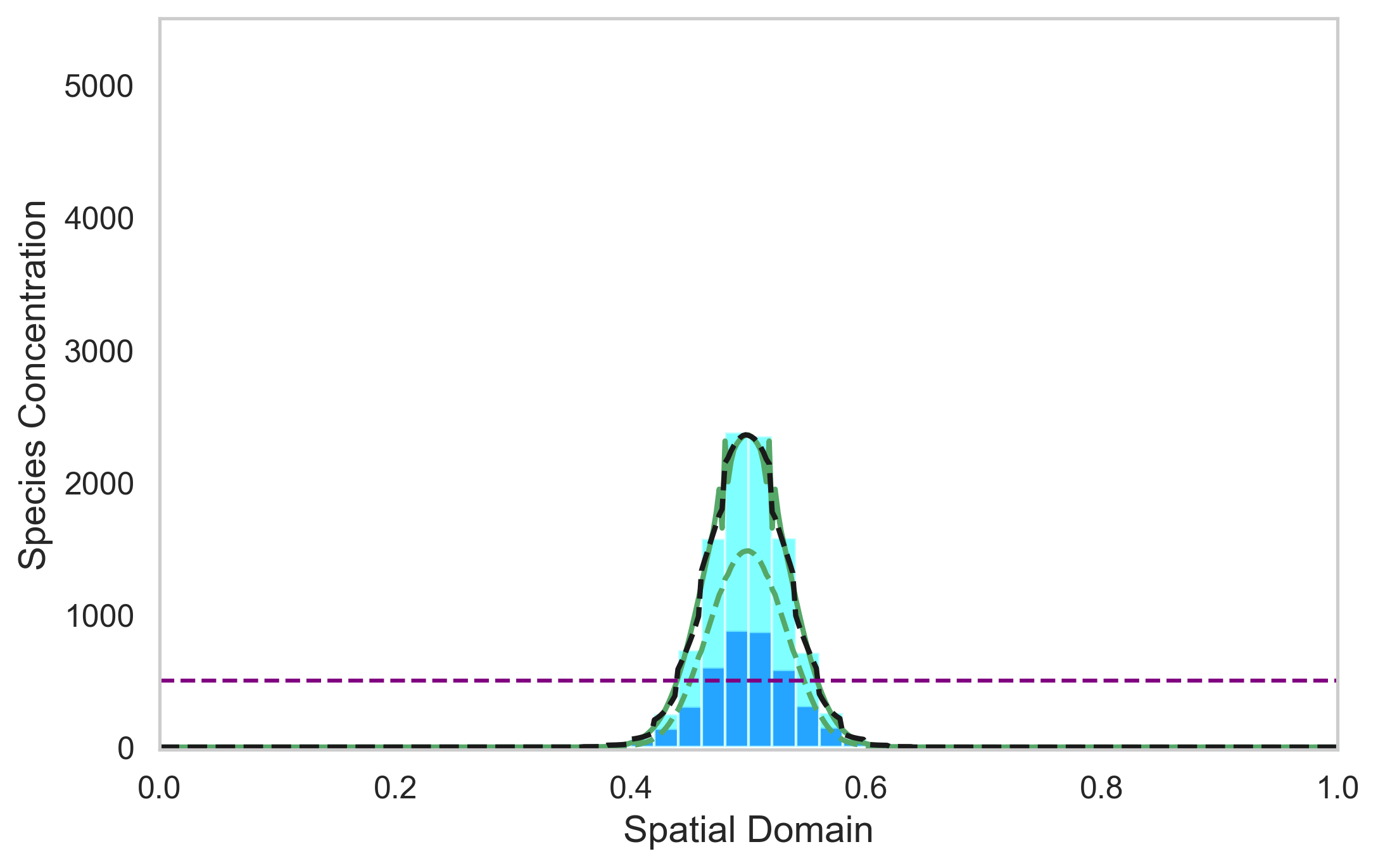}
        \caption{$t=0.5$}
        \label{fig:diff2}
    \end{subfigure}
    \begin{subfigure}{0.4\textwidth}
        \includegraphics[width=\linewidth]{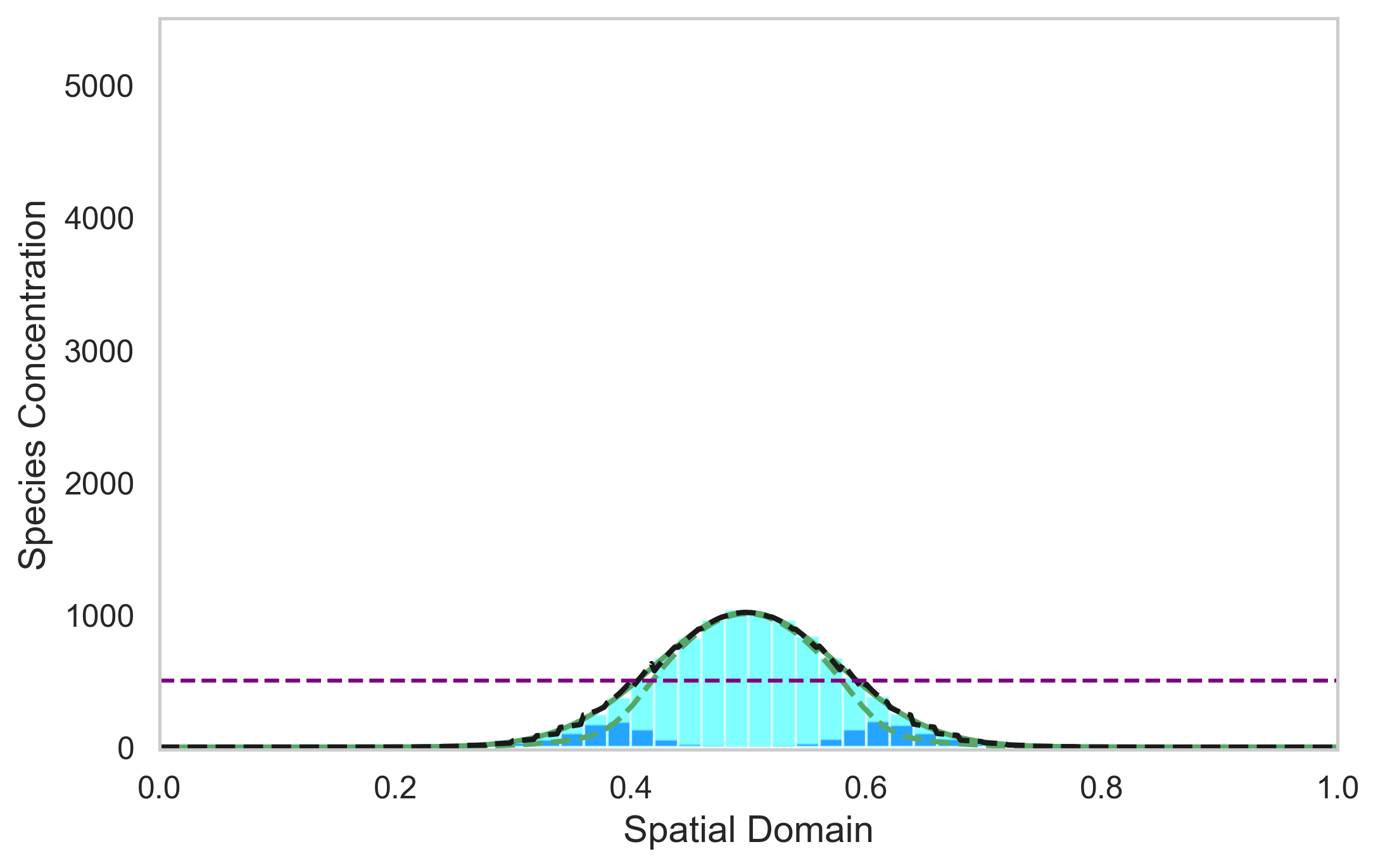}
        \caption{$t=3$}
        \label{fig:diff3}
    \end{subfigure}
    \begin{subfigure}{0.4\textwidth}
        \includegraphics[width=\linewidth]{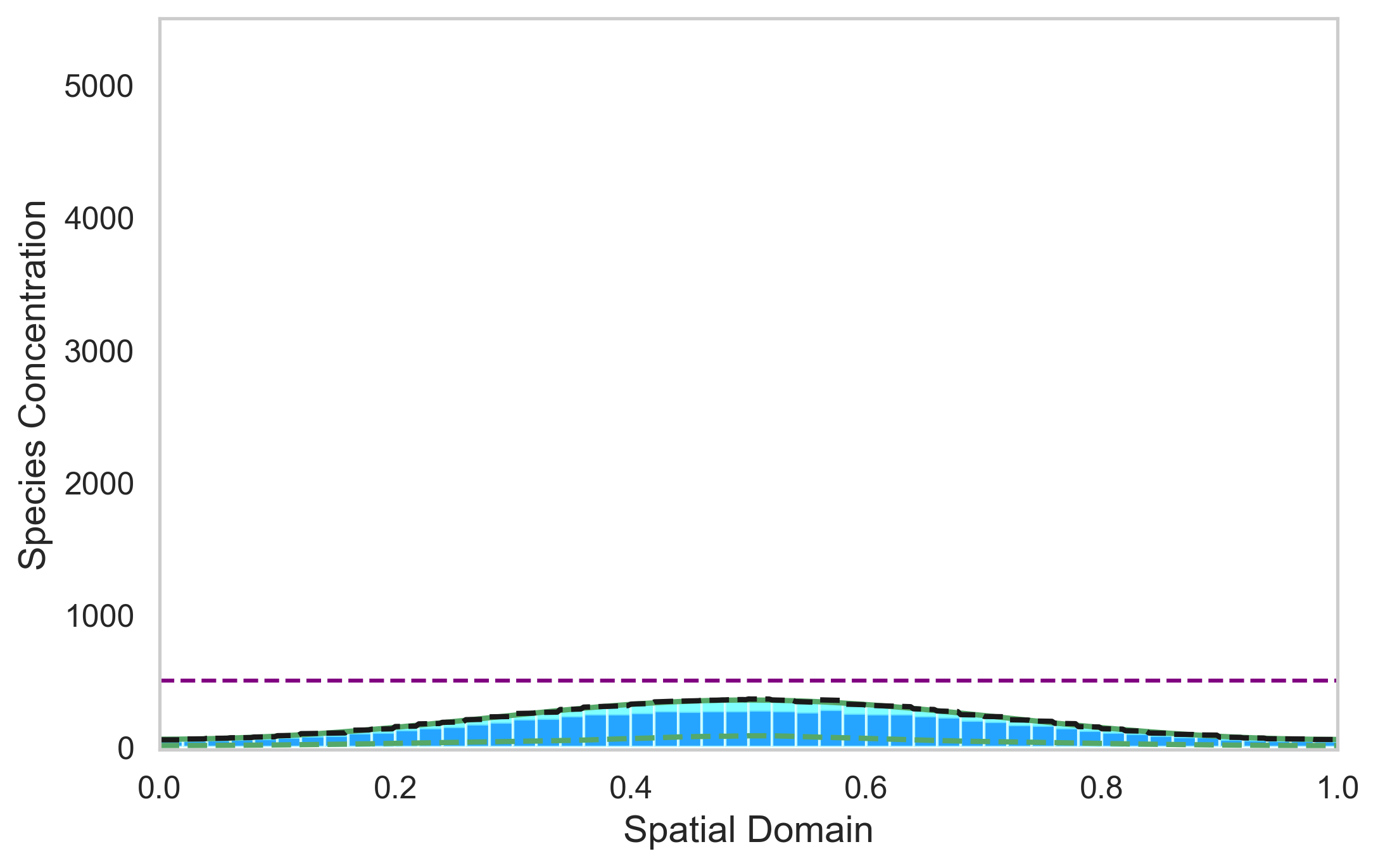}
        \caption{$t=25$}
        \label{fig:diff4}
    \end{subfigure}
    \caption{Diffusion from a top-hat initial condition, in which all the mass is initially in the discrete regime. The light blue histogram shows the results of the canonical stochastic diffusion model. The solid green line indicates the continuous PDE regime. The dark blue bars represent the discrete stochastic regime of the SRCM, while the dashed green line represents the mass in the PDE regime of the SRCM and the black dashed line represents the combined mass in the SRCM. Parameters: $\Omega = [0,1]$, $h = 0.02$, $\Delta t = 0.05$, $\Delta x = 2.5 \times 10^{-3}$, $\gamma = 2$, $\Theta = 500$, $D = 10^{-3}$. Results for the stochastic models are averaged over 500 independent repeats.}
    \label{fig:diffusion}
\end{figure}

In figure \ref{fig:diffusion} we plot $\bm{N}_H(t), \bm{N}_S(t)$, the ensemble averages over $500$ repeats of the hybrid and stochastic models, respectively, at four different time points of the simulation. We also plot the solution of the corresponding mean-field PDE \eqref{equation:mean_field_diffusion_PDE} at the same time points.

Figure \ref{fig:diffusion} demonstrates the spread of mass as it diffuses outwards from a top-hat initial condition (panel (\subref{fig:diff1})). The SRCM begins with all mass entirely in the stochastic regime. Conversion of mass to the continuous regime with rate $\gamma$ begins immediately and continues while the combined mass is above the conversion threshold. In panel (\subref{fig:diff2}) we can see the impact of the conversion, rapidly reducing the mass in the stochastic regime and converting it to continuous mass. By time $t=3$, almost all the mass has been converted to the continuous regime. However, as the mass diffuses outwards (see panel (\subref{fig:diff3})) the combined mass in the outer compartments sits below the threshold and mass  is either converted back to or remains as discrete mass. Finally, in panel (\subref{fig:diff4}) the combined mass is below threshold across the whole domain so the majority of the mass is converted back into the discrete regime. Note the good agreement between the canonical stochastic model (light blue bars) and the combined SRCM mass (dashed black line) (and indeed the mean-field PDE mass (solid green line)).

To analyse the error introduced by hybridisation, we compare the averaged profile of the hybrid method against the purely stochastic method. Separately, to understand the error between discrete and continuum approaches, we also compare the SSA against the deterministic PDE. To determine the level of self-error within the SSA, we compare the averaged profile of two independent SSA simulations. Let $\bm{N}_H(t)$ denote the ensemble average of the hybrid method, computed from $n = 1000$ independent repeats, each initiated from identically distributed initial conditions. Similarly, let $\bm{N}^1_S(t)$ and $\bm{N}^2_S(t)$ denote two independent ensemble averages of the SSA, each based on 1000 independent repeats. The deterministic PDE solution is denoted by $\bm{N}_P(t)$, converted to a pseudo-particle representation for comparison. We define $\hat{\bm{N}}_H(t)$, $\hat{\bm{N}}^1_S(t)$, $\hat{\bm{N}}^2_S(t)$, and $\hat{\bm{N}}_P(t)$ to be the normalised ensemble profiles, where normalisation is performed with respect to the total mass in the domain at each time point. The normalised histogram distance error between the hybrid and stochastic simulations is given by:

\begin{equation}
    \varepsilon_H(t) = \frac{1}{2} \sum_{k=1}^{K} \left| (\hat{N}_H)_k(t) - (\hat{N}^1_S)_k(t) \right|,
\end{equation}
and the self-error of the stochastic simulations is defined as:
\begin{equation}
    \varepsilon_S(t) = \frac{1}{2} \sum_{k=1}^{K} \left| (\hat{N}^1_S)_k(t) - (\hat{N}^2_S)_k(t) \right|.
\end{equation}
Similarly, the error between the hybrid and the PDE method is defined by:
\begin{equation}
    \varepsilon_P(t) = \frac{1}{2} \sum_{k=1}^{K} \left| (\hat{N}_S)_k(t) - (\hat{N}_P)_k(t) \right|.
\end{equation}

\begin{figure}[H]
    \centering
    \includegraphics[width=0.9\linewidth]{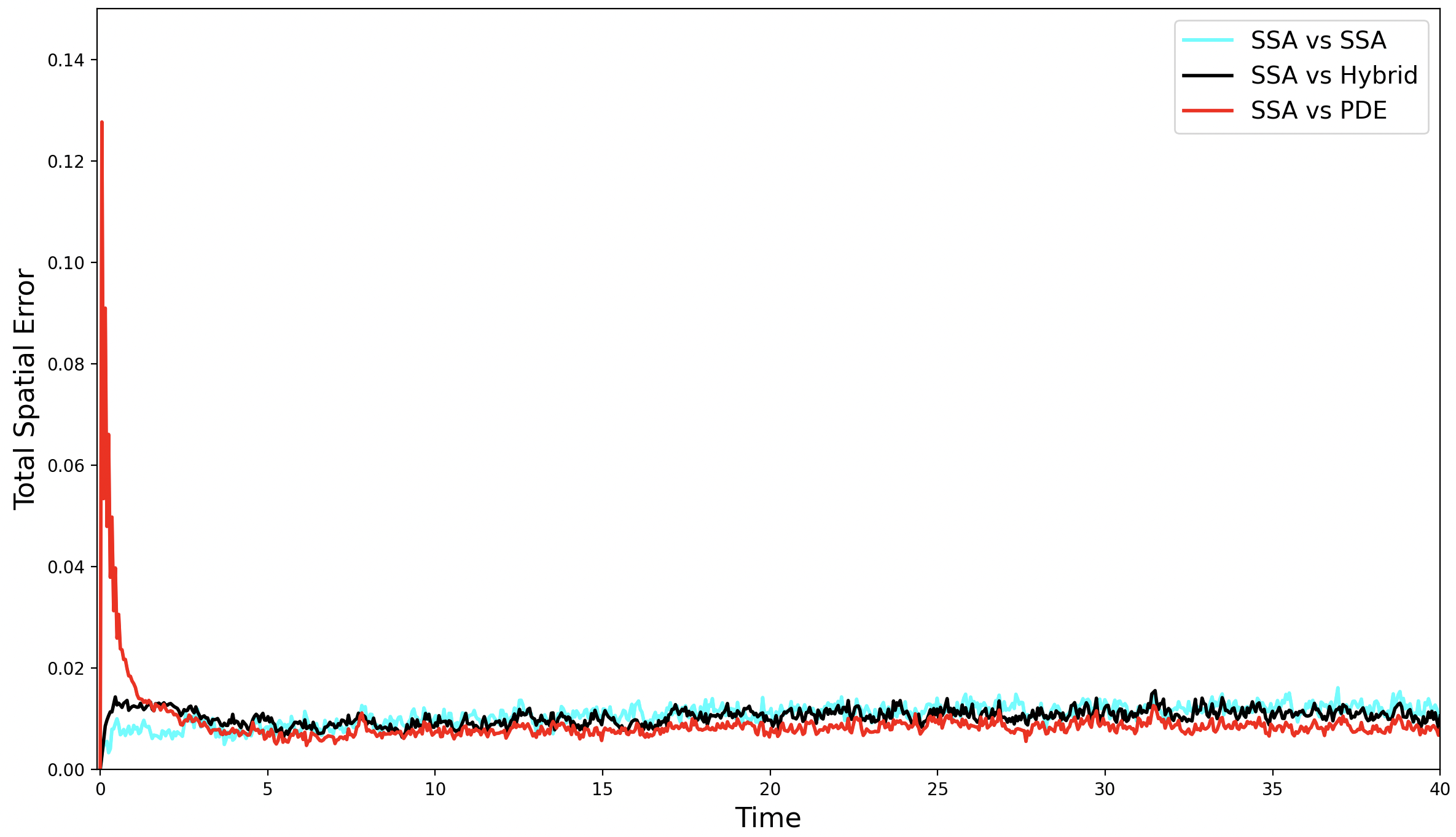}
    \caption{The evolution of the self-error ($\varepsilon_S$ - cyan line), the SRCM error $(\varepsilon_H$ - black line) and the PDE error $(\varepsilon_P$ - red line) over time for the simple diffusion test case.}
    \label{fig:diffusion error}
\end{figure}

We observe in figure \ref{fig:diffusion error} that the hybrid error is slightly elevated during the first few time steps of the simulation. This is likely due to the high rate of conversions between the discrete and continuous regimes in this early phase. The PDE model also exhibits an altered mass profile compared to the compartment-based canonical stochastic model, which is reflected in the large initial spike in the PDE error. When large particle masses are highly localised (as in the initial condition of figure \ref{fig:diffusion}), this discrepancy is likely to be more pronounced, potentially amplifying the PDE error.

As the simulation progresses, the hybrid error decreases and quickly comes to fluctuate at or below the self-error of the canonical stochastic simulation. Notably, the PDE model yields the lowest error relative to the hybrid. This is expected, since the PDE accurately captures the mean behaviour of the stochastic system and does not contribute intrinsic noise to the error. In contrast, the self-error between two ensemble stochastic averages leads to a higher relative error when compared to a single deterministic mean as both sets of simulations contribute their own intrinsic noise to the error.

This pure diffusion test case provides a useful benchmark for assessing the performance of the SRCM before introducing additional complexity in the form of zeroth-, first-, and second-order kinetic reactions.

\subsection{Case 2: Alternating growth and decay}

In this test case, we model a production-degradation reaction-diffusion system involving a single species, $\mathbf{A} = \{\text{A}\}$. The production and degradation processes are activated alternately over contiguous, non-overlapping time intervals. Specifically, let $T_1$ denote the set of time intervals during which production is active, and let $T_2$ denote the intervals during which degradation is active. These intervals are disjoint and together span the full simulation time, i.e., $T_1 \cup T_2 = [0, T]$ and $T_1 \cap T_2 = \emptyset$. We define the indicator function $\bm{1}_{T_1(t)}$ such that $\bm{1}_{T_1(t)} = 1$ if $t \in T_1$, and $0$ otherwise.

The production degradation system is captured by the following pair of reactions:
\begin{align}
    \text{r}_1 &: \text{A} \xrightarrow{\lambda_1 \bm{1}_{T_1(t)}} \emptyset, \\
    \text{r}_2 &: \emptyset \xrightarrow{\lambda_2 \bm{1}_{T_2(t)}} \text{A},
\end{align}
consistent across every compartment.
These are a first- and a zeroth-order reaction respectively, which means the reaction set in the SRCM is  
$\bar{\text{R}}^{\text{kin}} = [\bar{\text{r}}_1, \bar{\text{r}}_2]$ where:
\begin{align}
    \bar{\text{r}}_1 &: \text{D} \xrightarrow{\lambda_1 \bm{1}_{T_1}(t)} \emptyset, \\
    \bar{\text{r}}_2 &: \emptyset \xrightarrow{\lambda_2 \bm{1}_{T_2}(t)} \text{D}.
\end{align}
Consequently, the complete reaction system is given by $\bar{\text{R}}_k = [\text{R}^{\text{con}}, \text{R}^{\text{diff}}, \bar{\text{R}}^{\text{kin}}]$.

The corresponding PDE governing the concentration, $u(x,t)$, in the mean-field model is given by the following:
\begin{equation}
    \frac{\partial u}{\partial t} = D \frac{\partial^2 u}{\partial x^2} + \lambda_2 \bm{1}_{T_2}(t) - \lambda_1 \bm{1}_{T_1}(t) u, \quad x\in (0,L),\quad t\in(0,T].\label{equation:mean_field_on_off_PDE}
\end{equation}
As in test case 1, we implement zero-flux boundary conditions:
\begin{equation}
\frac{\partial u}{\partial x}\Big|_{x=0,L}=0.\nonumber
\end{equation}
Observe that if this were the PDE we used for the continuum representation in the SRCM, the zeroth-order production reaction would be implemented twice: once in the stochastic regime and once in the PDE. This would lead to an overproduction of mass. Since zeroth-order production is independent of concentration, including it in both regimes would result in a higher production rate than intended. In accordance with rule (C2), we implement production exclusively within the stochastic framework. This provides greater accuracy at low particle numbers, where discrete stochastic effects are most significant. Any mass that is produced can then transition into the PDE regime through standard conversion if the total mass is above the conversion threshold. As a result the PDE we solve in the SRCM is 
\begin{equation}
    \frac{\partial v}{\partial t} = D \frac{\partial^2 v}{\partial x^2} - \lambda_1 \bm{1}_{T_1}(t) v, \quad x\in (0,L),\quad t\in(0,T],\label{equation:hybrid_on_off_PDE}
\end{equation}
with the same zero-flux boundary conditions.

We set $K=20$, $\Omega = [0,1]$ and correspondingly, $h=0.05$. Setting the initial conditions for the SRCM as follows:
\[
    X_k(0) = 0 \quad \forall \ k \in \mathcal{K}, \quad v(x,0) = 1000 \ \forall \ x \in \Omega.
\]

The corresponding pure PDE has the following identical conditions, $u(x,0) = 1000$ for all $x \in \Omega$, while the canonical stochastic model is initialised with 
\[ X_k(0) = 50 \quad \forall \ k \in \mathcal{K}.
\]
Since we begin with a uniform initial condition we expect that our simulations will maintain uniformity throughout (at least in the ensemble average profiles). This absence of spatial heterogeneity allows us to analyse the total mass across the system, rather than focusing on local variations. We compare ensemble averages of the canonical stochastic model against the SRCM, the PDE solution and canonical model itself, as before.

\begin{figure}[H]
    \centering
    \includegraphics[width=0.9\linewidth]{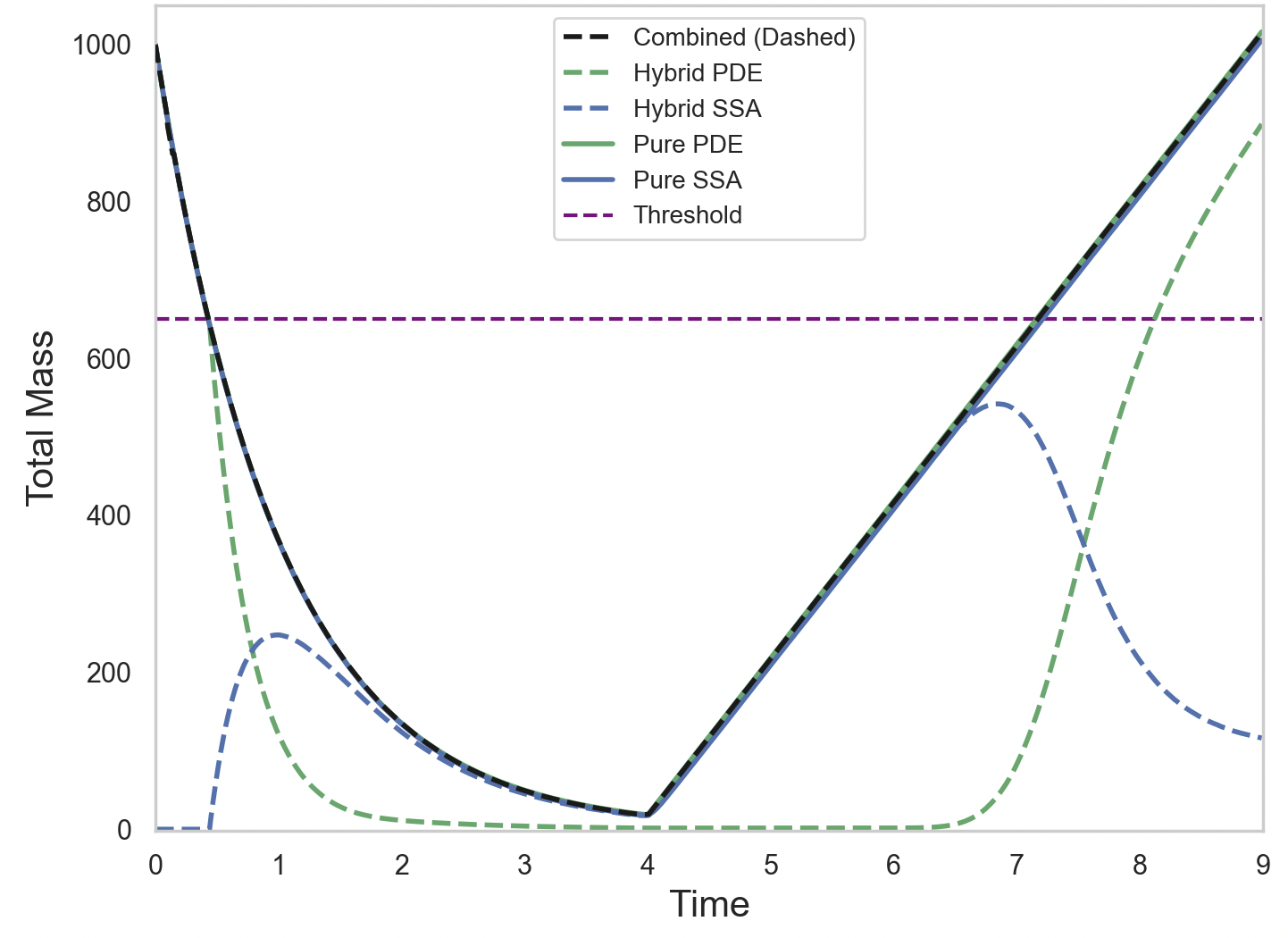}
    \caption{Total mass across the domain over time for the alternating growth and decay simulation.  The SRCM output includes: total hybrid mass (dashed black), PDE regime component (dashed green), and discrete stochastic component (dashed blue). Solid green and solid blue lines represent the pure PDE and canonical stochastic simulations, respectively. The conversion threshold is also shown. 
    Parameters are as follows: $\Omega = [0,1]$, $K=50$, $h = 0.02$, $P = 8$, $\Delta t = 0.05$, $\Delta x = 2.5 \times 10^{-3}$, $\gamma = 2$, $\Theta = 650$, $D = 10^{-3}$. Growth and decay rates are piecewise constant: $\lambda_1 = 1$ for $T_1 = [0,4]$ and $\lambda_2 = 200$ for $T_2 = (4,9]$. Results for the stochastic models are averaged over 1000 independent repeats.}
    \label{fig:Test case 1}
\end{figure}

Figure \ref{fig:Test case 1} illustrates the total mass of the SRCM, its constituent continuous and discrete components, as well as the total mass of the canonical stochastic simulation and the corresponding mean-field PDE. The discrete (dashed blue line) and continuous (dashed green line) regimes of the SRCM combine to form total mass (dashed black line). The combined mass in the SRCM agrees well with the mass of the canonical stochastic simulations (solid blue line) as well as with the PDE mass (solid green line); so much so that the PDE mass curve is obscured by the SRCM and canonical stochastic curves.

As the total number of particles falls below the threshold within the first time unit of the simulation, the continuous mass begins to be converted to discrete particles. The central part of the simulation is dominated by the stochastic regime before the production reaction is turned on and the degradation reaction is turned off. This allows the total mass to rise above the threshold for conversion, at which point stochastic mass is rapidly converted to continuous mass. It is worth noting that the discrete mass in the SRCM appears to decrease before the total mass reaches the conversion threshold. This is an artefact of averaging over a large number of repeats of the stochastic SRCM algorithm. Some of these repeats will reach threshold earlier than the average profile, while others will reach threshold later. When averaged over multiple repeats this gives the impression that the ensemble average mass drops before reaching the threshold, but in any individual repeat this will not be the case.

\begin{figure}[H]
    \centering
    \includegraphics[width=0.9\linewidth]{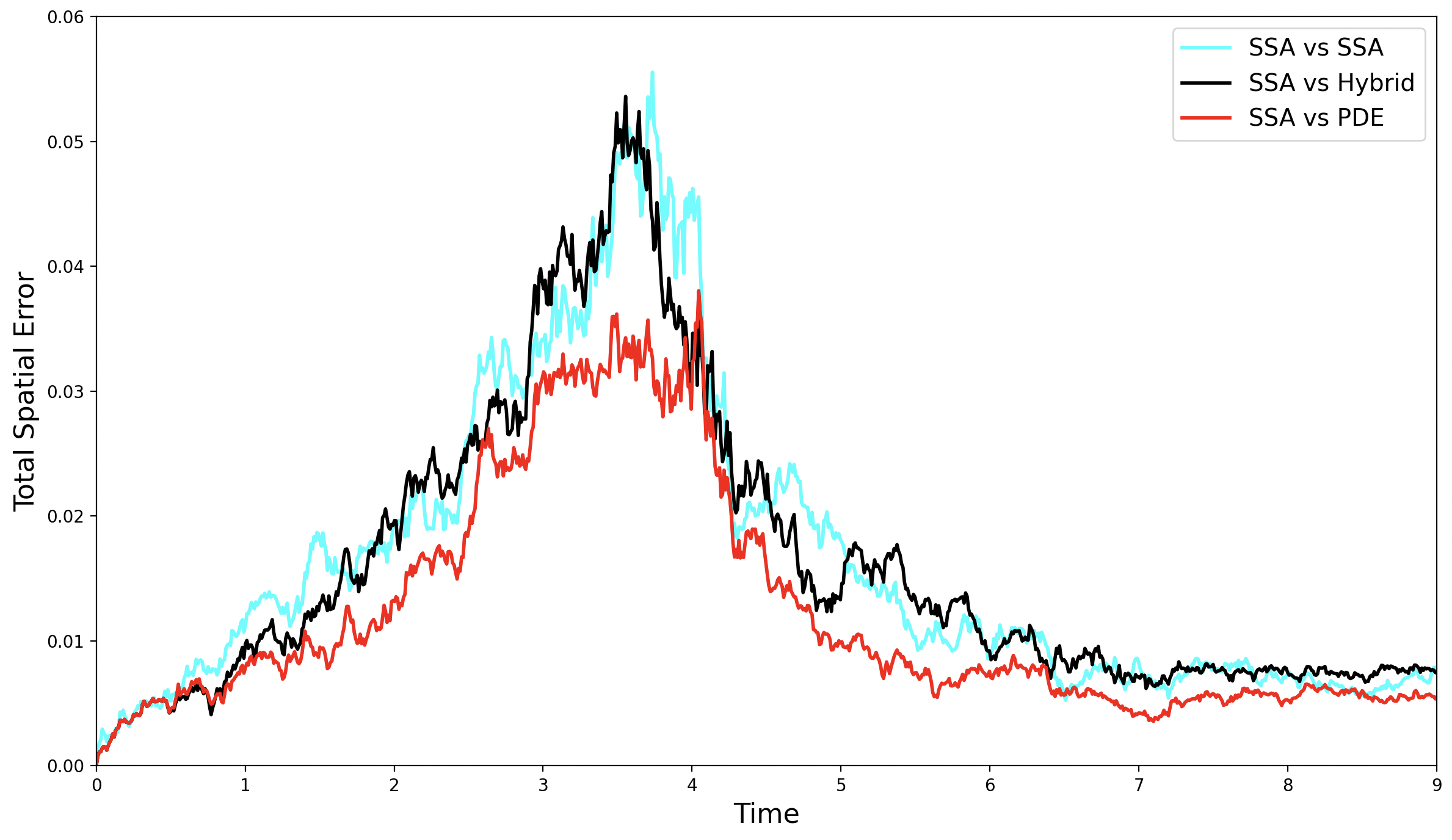}
    \caption{The evolution of the self-error ($\varepsilon_S$ - cyan line), SRCM error ($\varepsilon_H$ - black line) and PDE error ($\varepsilon_P$ - red line)  over time for the alternating growth and decay test problem.}
    \label{fig:Test case 1 error}
\end{figure}

In figure \ref{fig:Test case 1 error}, we show the stochastic self-error (cyan line), SRCM error (black line), and PDE error (red line), each measured relative to the canonical stochastic simulation. We can see that all three errors rise as the total mass in the system falls. This is a result of the fact that the fluctuations in the stochastic system have a bigger relative impact when the total mass in the system is smaller. The smallest error is observed between the PDE and the SSA. This is expected, as the PDE model provides an exact representation of the average behaviour of the stochastic system, and thus only one source of noise, the stochastic simulation, contributes to the discrepancy. In contrast, the self-error involves comparing two independent stochastic runs, each subject to its own noise, resulting in the largest relative error overall. Importantly, the SRCM error is comparable to the self-error suggesting the SRCM is doing a good job of replicating the canonical stochastic dynamics. 


\subsection{Case 3: Morphogen gradient} \label{Section:Morphogen}

In this test case, we model the formation of a morphogen gradient throughh a reaction-diffusion system in which there is uniform degradation of particles across the domain with production occurring at the left boundary, $x = 0$. As before, we consider just a single species, $\textbf{A} = \{ \text{A}\}$, this time with the following kinetic reactions within the canonical stochastic system:
\begin{equation}
    \begin{aligned}
        &\text{r}_1: \emptyset \xrightarrow{\lambda_{1,k}} \text{A}_k, \\
        &\text{r}_2: \text{A}_k \xrightarrow{\lambda_{2,k}} \emptyset.
    \end{aligned}
\end{equation}

So far we have assumed that these rates are consistent between compartments. In order to model the influx of particles into the domain in the stochastic representation, a zeroth-order production reaction will be implemented in the first compartment of the domain only. This is the first test case for which there are different reaction rates occurring in different compartments. Specifically we let, $\lambda_{1,1} = \lambda_1$ but set $\lambda_{k,1} = 0 \  \text{for} \ k \in \{2,\dots,K\}$.  In doing so, we switch off production in all compartments except the first to mimic an influx of particles over the left-hand boundary. However, we set $\lambda_{2,k}=\lambda_2$ to be the same across all compartments.

In the SRCM the reaction set is $\bar{\text{R}}^{\text{kin}} = \{\bar{r}_1,\bar{r}_2\}$ where:

\begin{equation}
    \begin{aligned}
        &\bar{\text{r}}_1: \emptyset \xrightarrow{\lambda_{1,k}} \text{D}_k, \\
        &\bar{\text{r}}_2: \text{D}_k \xrightarrow{\lambda_{2,k}} \emptyset,
    \end{aligned}
\end{equation}
 and $\lambda_{1,k}$ and $\lambda_{2,k}$ are as previously described.

The corresponding mean-field PDE model for the system is:

\begin{equation}
    \frac{\partial u}{\partial t} = D\frac{\partial^2 u}{\partial x^2}-\lambda_1u, \quad x\in (0,L),\quad t\in(0,T].\label{equation:mean_field_morphogen_gradient_PDE}
\end{equation}
The system has the following boundary conditions corresponding to production occurring on the left boundary and a zero-flux right-hand boundary:

\begin{equation}
    D\frac{\partial u}{\partial x}\Big|_{x=0}= -\lambda_2, \quad \frac{\partial u}{\partial x}\Big|_{x=L}=0.
\end{equation}

As with the top-hat test case, the solution of the PDE part of the SRCM, $v(x,t)$ evolves according to the same PDE as $u(x,t)$, but with zero-flux boundary conditions at \textit{both} ends:
\[
\frac{\partial v}{\partial x}\Big|_{x=0,L}=0.
\]
We set $K=40$, $\Omega = [0,1]$ and correspondingly $h=0.025$. The initial conditions for the SRCM are:
\[
X_k(0) = 1 \ \forall\ \ k \in \mathcal{K}, \quad v(x,0) = 0.
\]
The corresponding initial condition for the pure PDE is $u(x,0) = 40$ for all $x \in \Omega$, while the canonical stochastic model is initialised with \[
X_k(0) = 1 \ \forall\ \ k \in \mathcal{K}.
\] 
In the SRCM we let mass enter the system through the left boundary the stochastic regime as described above and in accordance with rule (C2).
\begin{figure}[H]
    \centering
    \begin{subfigure}{0.45\textwidth}
        \includegraphics[width=\linewidth]{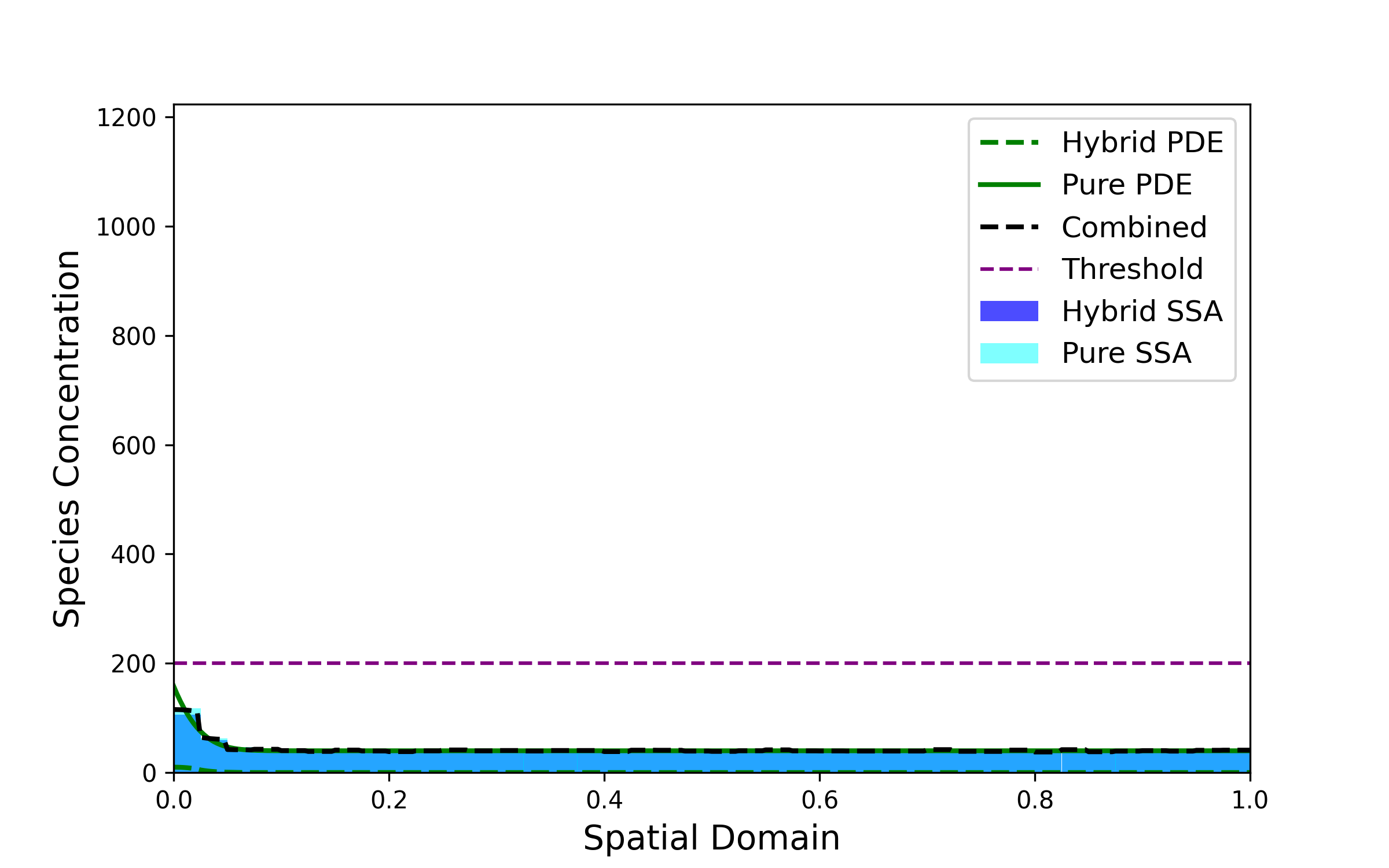}
        \caption{$t=0.5$}
        \label{fig:diff_t0}
    \end{subfigure}
    \begin{subfigure}{0.45\textwidth}
        \includegraphics[width=\linewidth]{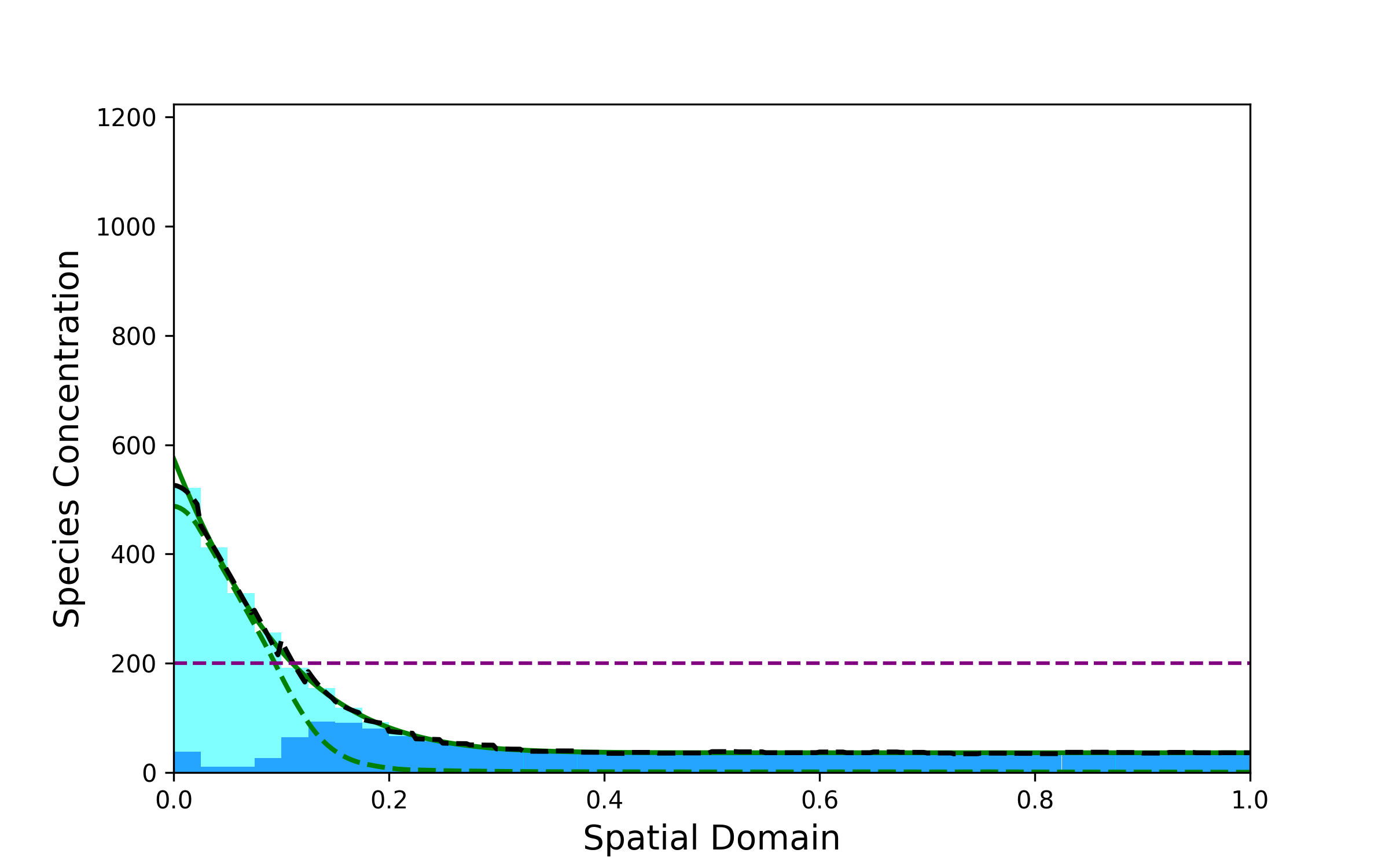}
        \caption{$t=10$}
        \label{fig:diff_t05}
    \end{subfigure}
    \caption{Morphogen gradient test problem, with all mass initially in the discrete regime. The SRCM consists of: the discrete stochastic component (dark blue histogram), the PDE component (dashed green line) which are combined to form the total mass (dashed black line). The canonical stochastic solution is shown in cyan, and the pure PDE solution in solid green.  Parameters: $\Omega = [0,1]$, $h = 0.025$, $\Delta t = 0.005$, $\Delta x = 3.125 \times 10^{-3}$, $\gamma = 4$, $\Theta = 200$, $D = 10^{-3}$, $\lambda_1 = 5$, $\lambda_2 = 10^{-2}$. Results for the stochastic models are averaged over 1000 independent repeats.}
    \label{fig: Morphogen}
\end{figure}
Figure \ref{fig: Morphogen} demonstrates that mass is produced at the left hand boundary and diffuses across the domain while also degrading to form an exponentially decaying profile. Once the concentration surpasses the conversion threshold, the majority of the mass in those regions is converted to the continuous PDE regime. 
Notably, we observe an elevated number of discrete particles in the first compartment relative to elsewhere in the domain, which is a result of the production of mass at the boundary being implemented in the discrete regime before it transitions into the PDE regime. As we move rightwards into the domain, the discrete mass increases due to C-D conversion reactions as the combined profile drops below threshold, before eventually entering the region almost entirely governed by the discrete regime, where the total concentration falls sits well below the threshold.

One potential issue with this implementation of the SRCM is the high number of conversion reactions occurring within the first compartment as a result of the influx being implemented in the stochastic regime according to rule (C2). We do not expect this to be a substantial computational cost in comparison to the potential savings made by employing the PDE instead of simulating large numbers of discrete particles. Nevertheless, one potential solution might be to produce mass stochastically when the combined mass is below threshold and then to produce via the PDE when above threshold. This would, however, add some overheads to the code and so is perhaps of questionable benefit.
\begin{figure}[H]
    \centering
    \includegraphics[width=0.9\linewidth]{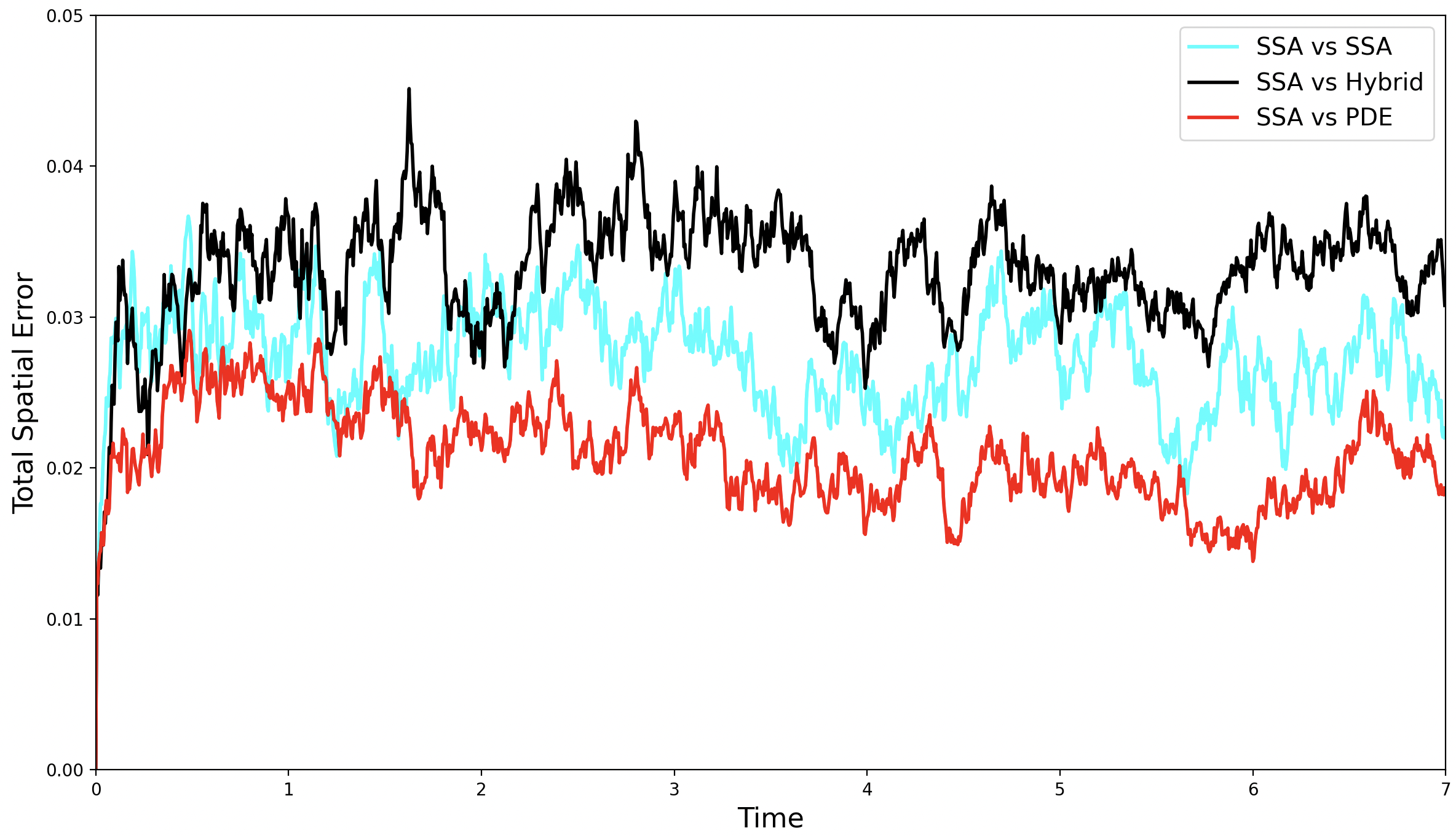}
    \caption{The evolution of the self-error ($\varepsilon_S$ - cyan line), SRCM error ($\varepsilon_H$ - black line) and PDE error ($\varepsilon_P$ - red line)  over time for the morphogen gradient test problem.}
    \label{fig:Mophgradient Error}
\end{figure}

Figure \ref{fig:Mophgradient Error} shows both the stochastic self-error, SRCM error and PDE error computed over 1000 simulations. While the errors are of the same order of magnitude, we observe, as before, that the hybrid error is of a similar order of magnitude to the stochastic self-error.
 

\subsection{Case 4: FKPP travelling wave} \label{Section:FKPP}

The final test problem we will consider is the FKPP (Fisher-Kolmogorov-Petrovsky-Piskunov) \citep{kolmogorov1937study,fisher1937wave} travelling wave system. Travelling waves are common phenomena seen in biological systems at different scales, from animal migration \citep{Simpson_McCue_2024} to cell migration \citep{Vittadello_McCue_Gunasingh_Haass_Simpson_2018} and so represent an important and well-used candidate for hybrid modelling.
The FKPP reaction-diffusion model contains a non-linear logistic growth term paired with diffusion. This is a thorough test case for the SRCM, as the front’s stochasticity critically influences wave speed \citep{Moro_2004,Harrison_Yates_2016}.

The FKPP equation makes for an excellent final test problem for the SRCM since it is well known that the typical stochastic model used to simulate FKPP-like dynamics (see reactions \eqref{equation:FKPP_reaction_1}-\eqref{equation:FKPP_reaction_2}) differs quantitatively from the moment-closed PDE version of the model (see equation \eqref{equation:mean_field_fisher_PDE}). Successfully replicating the wave speed of the stochastic analogue of the FKPP PDE will demonstrate that the SRCM is accurately capturing the important features of the full canonical stochastic model.

Again consider a single species, $\textbf{A}=\{\text{A}\}$. In the canonical stochastic system we consider the following set of kinetic reactions $\text{R}^{\text{kin}} = \{\text{r}_1,\text{r}_2\}$:  

    \begin{align}
        \text{r}_1: \text{A}_k \xrightarrow{\lambda_{1,k}} 2\text{A}_k,
        \label{equation:FKPP_reaction_1}\\ 
        \text{r}_2: 2\text{A}_k 
        \xrightarrow{\lambda_{2,k}} \text{A}_k.
        \label{equation:FKPP_reaction_2}
    \end{align}

We define the extended reaction network $\bar{\text{R}} = \{\text{R}^{\text{con}}, \text{R}^{\text{diff}}, \bar{\text{R}}^{\text{kin}}_D\}$, in which the extended kinetic reaction set is $\bar{\text{R}}^{\text{kin}}_D = \{\bar{r}_1,\bar{r}_2,\bar{r}_3, \bar{r}_4\}$ and corresponding propensities $\bar{\bm{\alpha}}_k=\{\alpha_{1,k},\alpha_{2,k},\alpha_{3,k}, \alpha_{4,k}\}$:

\begin{equation}
        \begin{aligned}
            &\bar{\text{r}}_1: \text{D}_k \xrightarrow{\lambda_{1,k}} 2\text{D}_k, && \alpha_{1,k} = \lambda_1 X_k,\\
            &\bar{\text{r}}_2: 2\text{D}_k \xrightarrow{\lambda_{2,k}} \text{D}_k, && \alpha_{2,k}=\lambda_2 \frac{X_k(X_K-1)}{h}, \\
            &\bar{\text{r}}_3: \text{D}_k+\text{C}_k \xrightarrow{\lambda_{2,k}} \text{C}_k, && \alpha_{3,k}=\lambda_2 \frac{X_k Y_k}{h}, \\
            &\bar{\text{r}}_4: \text{C}_k+\text{D}_k \xrightarrow{\lambda_{2,k}} \text{C}_k, && \alpha_{4,k}=\lambda_2 \frac{X_k Y_k}{h}. \\
        \end{aligned}
\end{equation}

Note that reactions \( \bar{r}_3 \) and \( \bar{r}_4 \) represent the same bimolecular interaction between \( \text{D}_k \) and \( \text{C}_k \), but are listed separately to account for the two distinct ways the molecules can be selected. The corresponding moment-closed PDE is as follows:
\begin{equation}
\frac{\partial u}{\partial t} = D \frac{\partial^2 u}{\partial x^2} + \lambda_1 u - \lambda_2 u^2, \quad x\in (0,L),\quad t\in(0,T].\label{equation:mean_field_fisher_PDE}
\end{equation}
As in test cases 1 and 2, we implement zero-flux boundary conditions:
\begin{equation}
\frac{\partial u}{\partial x}\Big|_{x=0,L}=0,\nonumber
\end{equation}
although we note that the boundary condition at $x=L$ has little effect since large amounts of mass do not reach the right-hand boundary during the course of our simulations. For this test case, we again ensure that the solution of the PDE component of the SRCM, $v(x,t)$, obeys the same dynamics and boundary conditions as the full PDE solution, $u(x,t)$. For the FKPP test case set $K=50$, $\Omega = [0,5]$ and consequently, $h=0.1$. We specify the initial conditions for the SRCM as follows:
\[
X_k(0) = 
\begin{cases}
    10, \quad k = 1,\\
    0, \quad k \in \mathcal{K} \setminus{1},
\end{cases} \quad v(x,0) = \mathbf{0}.
\]
The corresponding pure PDE initial conditions is $u(x,0) = 100\cdot\mathbf{1}_{\Omega_1}$, while the canonical stochastic model is initialised with \[
X_k(0) = 
\begin{cases}
    10, \quad k = 1,\\
    0, \quad k \in \mathcal{K} \setminus{1}.
\end{cases}
\]

\begin{figure}[H]
    \centering
    \begin{subfigure}{0.44\textwidth}
        \centering
        \includegraphics[width=\linewidth]{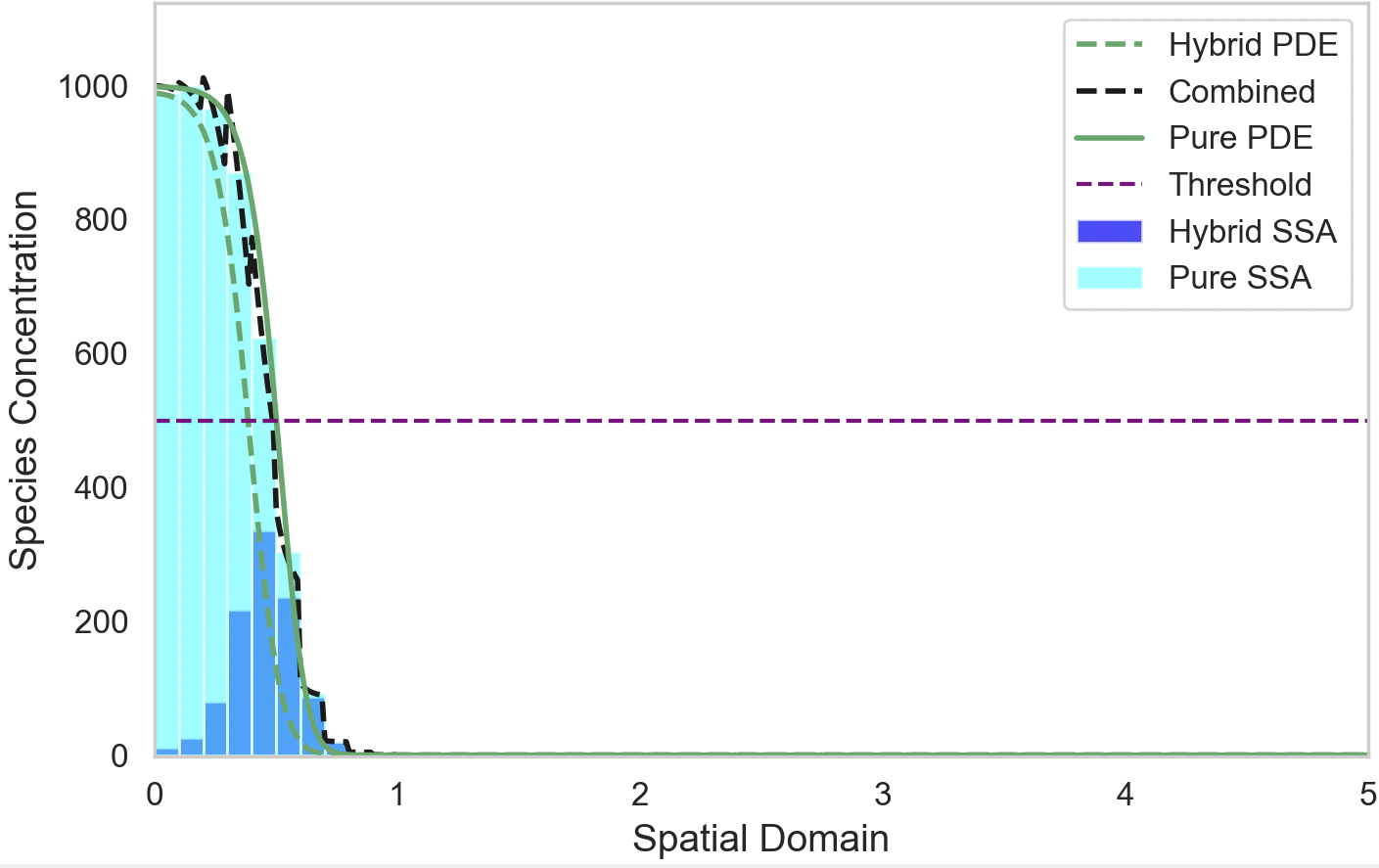}
        \caption{$t=0.5$}
        \label{fig:fig1}
    \end{subfigure}
    \begin{subfigure}{0.45\textwidth}
        \centering
        \includegraphics[width=\linewidth]{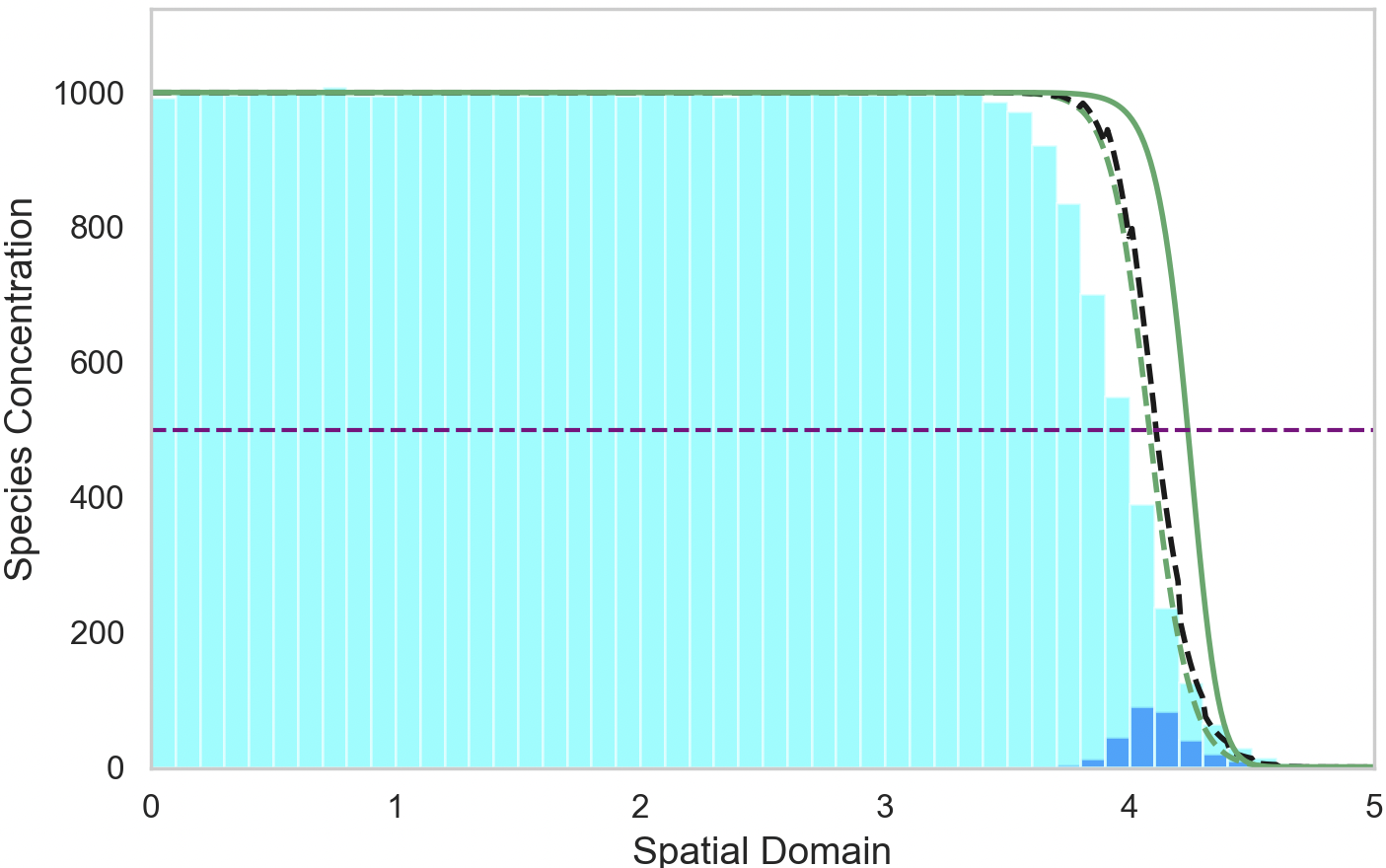}
        \caption{$t=7$}
        \label{fig:fig2}
    \end{subfigure}
    \caption{Solutions of the FKPP test problem, with all mass initially in the discrete regime. Figure descriptions are as in figure \ref{fig: Morphogen}. Parameters are as follows: $\Omega = [0,5]$, $h = 0.1$, $P = 8$, $\Delta x=1.25 \times 10^{-2}$, $T = 10$, $\Delta t = 0.005$, $\Theta = 500$, $\gamma = 1$, $\lambda_1 = 10$, $\lambda_2 = 0.01$, $D = 10^{-2}$. Results for the stochastic models are averaged over 1000 independent repeats.}
    \label{fig:naive_FKPP}
\end{figure}

Figure \ref{fig:naive_FKPP} shows the propagation of a travelling wave in the FKPP system. As expected, the concentration grows toward the stable steady state $\lambda_1/\lambda_2$ and the wave moves across the domain. In the hybrid implementation, mass is only modelled as discrete stochastic particles  towards the leading edge of the wave. Although the hybrid model reproduces a travelling wave, its profile and speed differs from the fully stochastic model which in turn differs from PDE solution.

\begin{figure}[H]
    \centering
    \begin{subfigure}[t]{0.48\textwidth}
        \centering
        \includegraphics[width=\linewidth]{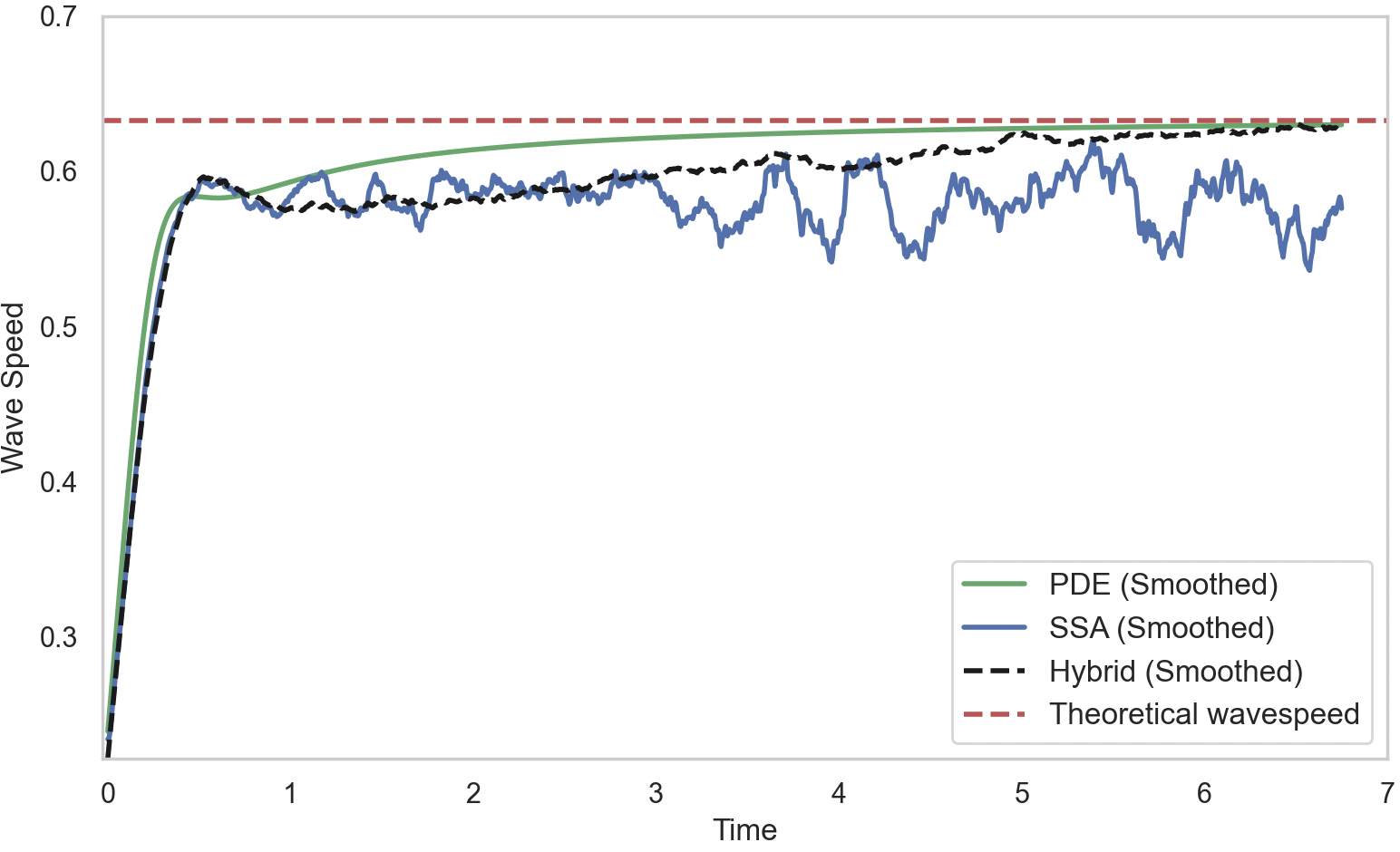}
        \caption{}
        \label{fig:naive_wavespeed}
    \end{subfigure}
    \hfill
    \begin{subfigure}[t]{0.48\textwidth}
        \centering
        \includegraphics[width=\linewidth]{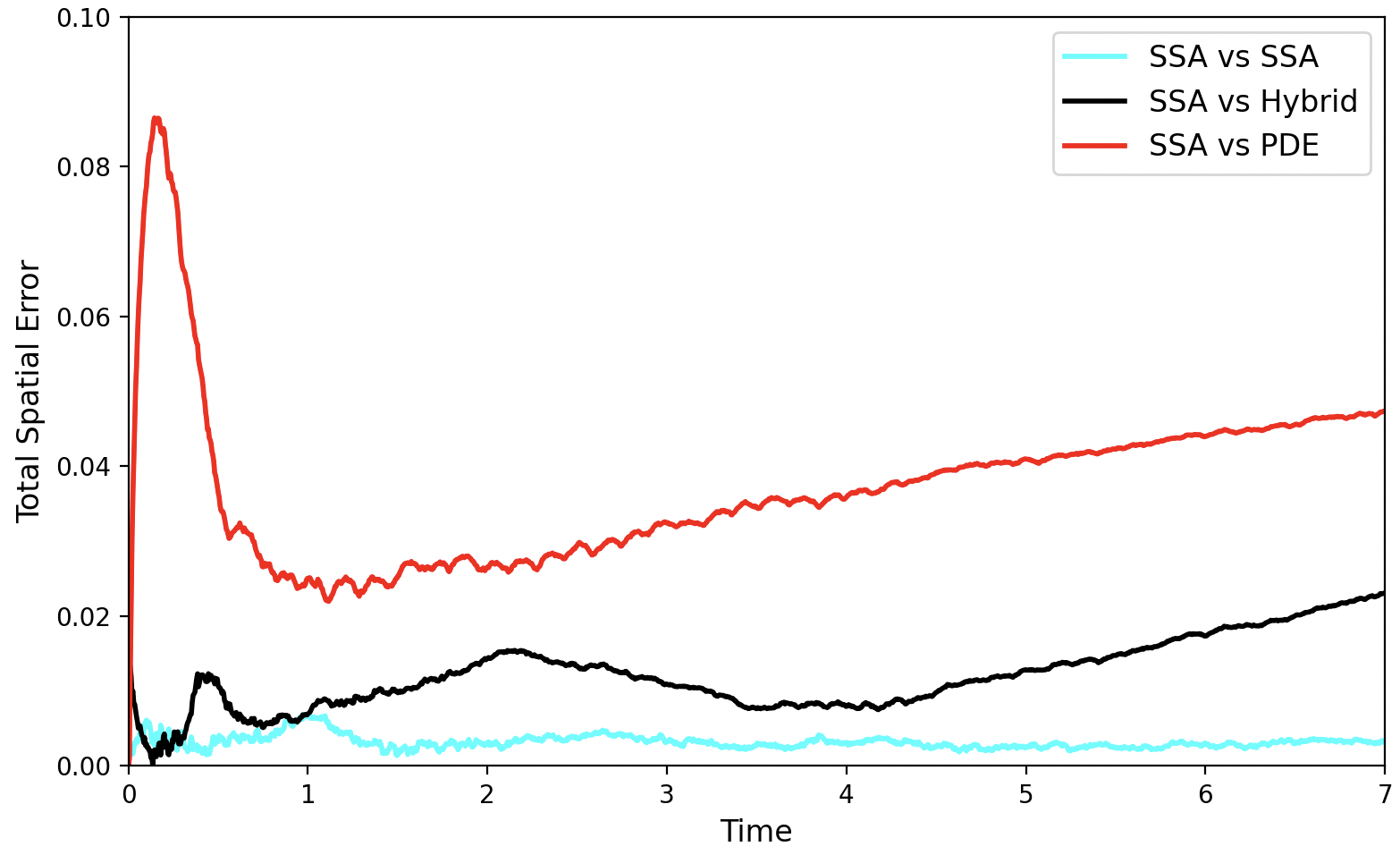}
        \caption{}
        \label{fig:fkpp naive error}
    \end{subfigure}
    \caption{Comparison of (a) wave speed and (b) histogram distance errors errors across profiles averaged over 1000 repeats (stochastic models). Parameters are as in figure \ref{fig:naive_FKPP}.}
    \label{fig:naive_wavespeed-error}
\end{figure}

To better understand this discrepancy, we analyse both the wave speed and spatial error between the SRCM and canonical stochastic models. The PDE travelling wave solution has a theoretical minimal wavespeed of $2\sqrt{D\lambda_1}$, which the numerical simulation approaches as the wave front stabilises (see figure \ref{fig:naive_wavespeed-error} (\subref{fig:naive_wavespeed}))\citep{Murray_2008}. The stochastic model, with a finite number of particles, converges to a wavespeed that is slower than the PDE.

Determining the wavefront position and speed directly is complicated by stochasticity. To overcome this, we adopt the approach of \citet{Robinson2014}, by estimating the wave speed from the rate of change in total system mass (this method is only truly valid when the shape of the wavefront does not change). Specifically, given time points $t_1$ and $t_2$, the wave speed is approximated by:

\begin{equation}
\hat{c} = \frac{N(t_2)-N(t_1)}{t_2-t_1}\cdot\frac{\lambda_1}{\lambda_2},
\end{equation}
where $N(t)$ is the total mass of species $\text{A}$ in the system at time $t$. Figure \ref{fig:naive_wavespeed-error} (\subref{fig:naive_wavespeed}) compares the estimated wave speeds of the PDE, the SSA, and  the SRCM models, alongside the theoretical minimum wave speed.

In figure \ref{fig:naive_wavespeed-error} (\subref{fig:naive_wavespeed}), we see that the smoothed travelling wave speed of the SRCM initially matches the stochastic simulation but gradually accelerates toward the PDE wave speed. To reduce the high-frequency noise inherent in stochastic simulations, we apply a moving average over a 50-time-point window. This smoothing reveals the underlying trend in wave propagation by dampening local fluctuations. The resulting divergence from the stochastic simulation is highlighted in the error plot in figure \ref{fig:naive_wavespeed-error} (\subref{fig:fkpp naive error}), where the SRCM increasingly deviates over time.

The convergence of the SRCM wavespeed to that of the PDE indicates a problem with the mass generation at the wavefront (where the wave speed is determined in this pulled front). According to rule (C2) mass is produced in the PDE regime according to the first-order production reaction. This mass can then diffuse to the wavefront, in sufficiently low concentrations that it cannot be converted into discrete mass, and where it is further amplified by first-orderproduction in the PDE regime. Effectively, this means the wave front is dominated by PDE dynamics rather than the discrete stochastic dynamics that should be dominating in order to simulate the correct wave speed. 

To correct this, production should be restricted to the stochastic regime ensuring that dynamics at the front remain consistent with the stochastic implementation for this pulled wavefront. However, there is a fine balance between efficiency and accuracy, so we specify that the SRCM should produce mass via the stochastic regime if total mass is below threshold and in the continuous regime if above threshold. This means that production will occur in the continuous regime   behind the wavefront, while production at the wavefront will occur exclusively in the stochastic regime.
We implement this change to the SRCM and demonstrate that it produces travelling waves with the correct wave speed in the following section.

\subsubsection{Adaptive production switching}

 We propose an adaptive mechanism to dynamically allocate production between the discrete and continuum regimes based on a combined threshold criterion. This ensures that stochastic effects are captured at the wavefront while maintaining computational efficiency behind the wave where stochasticity is less important, balancing efficiency and accuracy.

 Recall that $N_k(t) := X_k(t)+Y_k(t)$ where $Y_k(t) = \int_{\Omega_k} v(x,t)\mbox{d}x$, for $k \in \{1,\dots,K\}$. For each compartment $\Omega_k$, we would like to specify that if $N_k(t) \geq \Theta$ then the first-order production reaction occurs solely within the continuous PDE regime, otherwise it will occur solely within the stochastic discrete regime. This can be done using a vector of indicator functions $\bm{\beta}(t) \in \{0,1\}^K$, such that $\beta_k(t) = \bm{1}_{N_k(t)\geq \Theta}$. In the extended network, $\bar{\mathcal{N}}$, we adapt the propensity function of reaction $\bar{\text{r}}_1$ as follows:

 \[
 \alpha_{k,1}(t)=\lambda_1N_k(t)(1-\beta_k(t)).
 \]
Let 
\begin{equation}
\bar{u}(x,t)=\sum_{k=1}^K \, (v(x,t)+X_k(t)/h)\cdot\mathbf{1}_{\Omega_k}(x),
\end{equation}
define the entire combined mass at position $x$ at time $t$ (from both regimes) within the SRCM. The corresponding PDE for the continuous component of the SRCM is given by the following:
\begin{equation}
\frac{\partial v}{\partial t} = D\frac{\partial^2 v}{\partial x^2} + \lambda_1 \left[ \sum_{k=1}^K \beta_k(t)\, \mathbf{1}_{\Omega_k}(x) \right] \cdot\bar{u}(x, t) - \lambda_2 v^2\quad x \in (0, L),\ t \in (0, T].
\end{equation}

To reflect the adaptive nature of the SRCM, we modify the PDE so that production terms are only active in compartments in which the total mass concentration exceeds the threshold, \(\Theta\). In practice, this means that if the mass concentration in compartment \(k\) falls below \(\Theta\), the corresponding source term in the PDE is turned off. This ensures that PDE production is suppressed in regions governed primarily by the discrete regime and only applied where a continuous approximation is valid.

\begin{figure}[H]
    \centering
    \begin{subfigure}{0.45\textwidth}
        \centering
        \includegraphics[width=\linewidth]{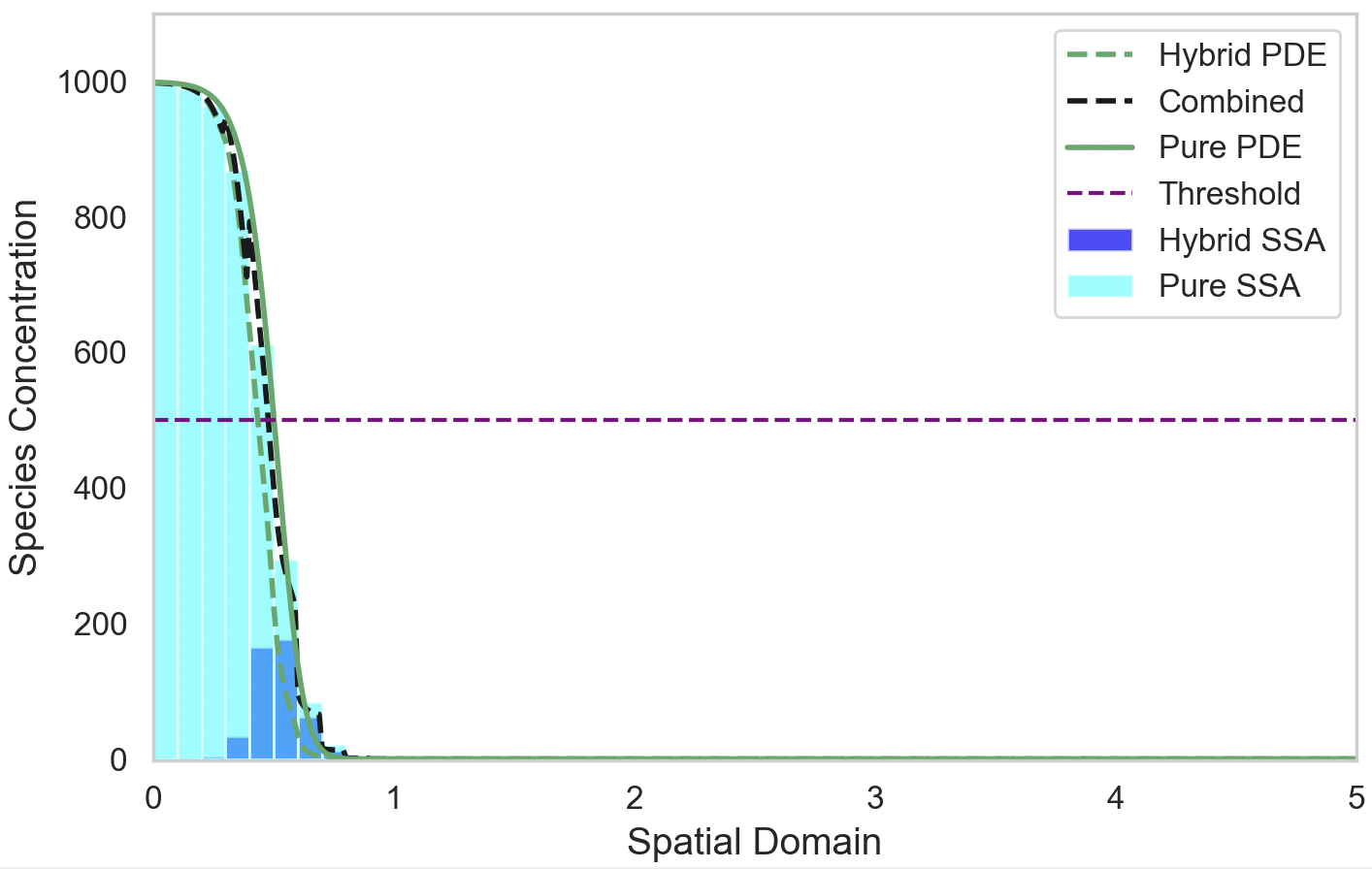}
        \caption{$t=0.5$}
        \label{fig:fig1_FKPP}
    \end{subfigure}
    \begin{subfigure}{0.45\textwidth}
        \centering
        \includegraphics[width=\linewidth]{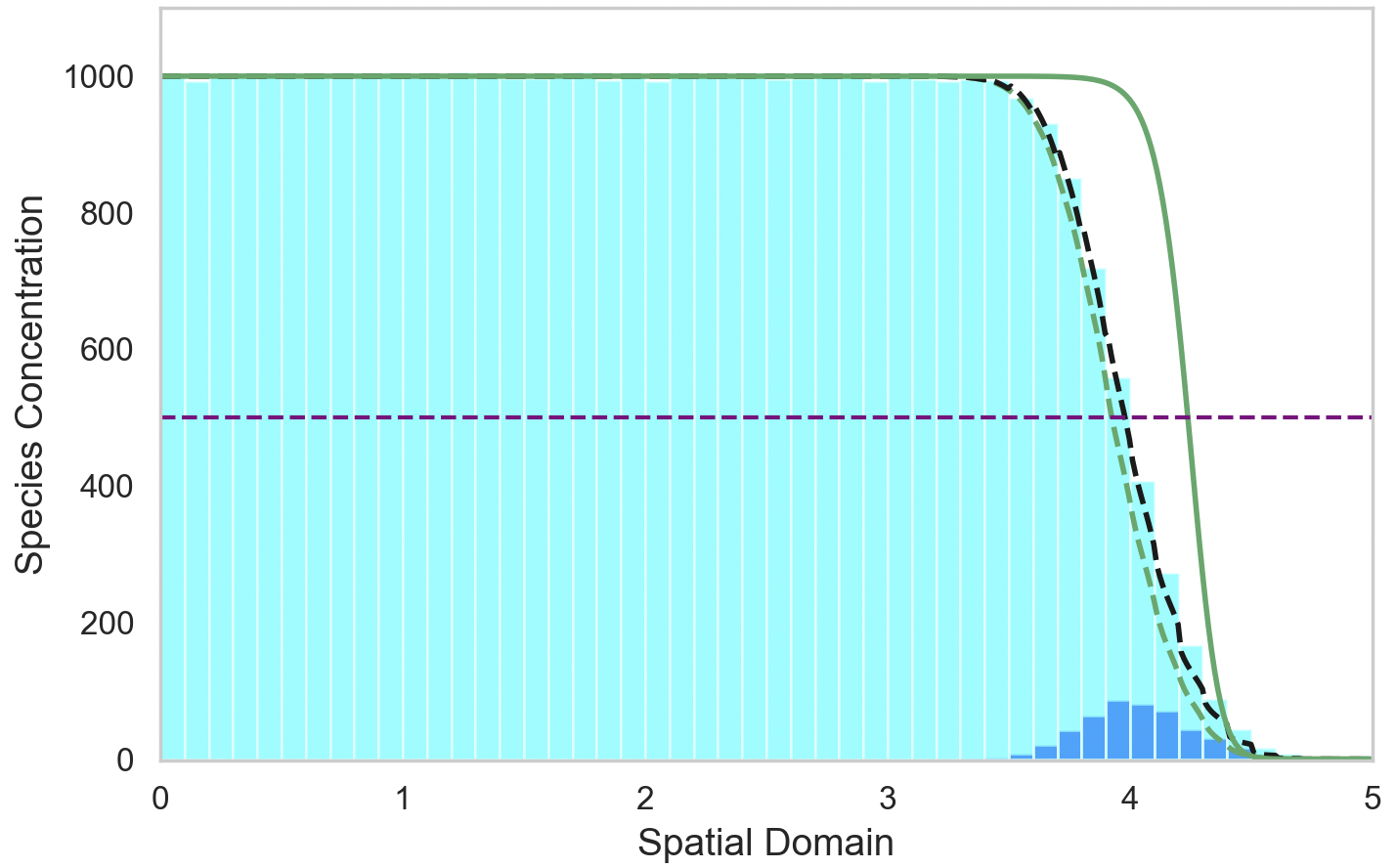}
        \caption{$t=7$}
        \label{fig:fig2_FKPP}
    \end{subfigure}
    \caption{Solutions of the FKPP test problem using the SRCM with adaptive production. Figure descriptions are as in figure \ref{fig: Morphogen}.   Parameters are as follows: $\Omega = [0,5]$, $h = 0.1$, $P = 8$, $\Delta x=1.25 \times 10^{-2}$, $T = 10$, $\Delta t = 0.005$, $\Theta = 500$, $\gamma = 1$, $\lambda_1 = 10$, $\lambda_2 = 0.01$, $D = 10^{-2}$. Results for the stochastic models are averaged over 1000 repeats.}
    \label{fig:FKPP}
\end{figure}

This adaption ensures that mass is produced via either the PDE or the stochastic regime depending on the total mass within each compartment (as opposed to production being constitutively turned on for both regimes). This adaptation does incur a small computational overhead associated with adding the discrete and continuous mass at each PDE timestep to find the total. However, not having to simulate stochastic production within each compartment should offset some if not all of the extra expense. In figure \ref{fig:FKPP} we plot the profiles of the adapted SRCM (alongside the reproduced solution profiles of the canonical stochastic system and the PDE) for the FKPP model with the same parameters in as figure \ref{fig:naive_FKPP}.

\begin{figure}[H]
    \centering
    \begin{subfigure}[t]{0.48\textwidth}
        \centering
        \includegraphics[width=\linewidth]{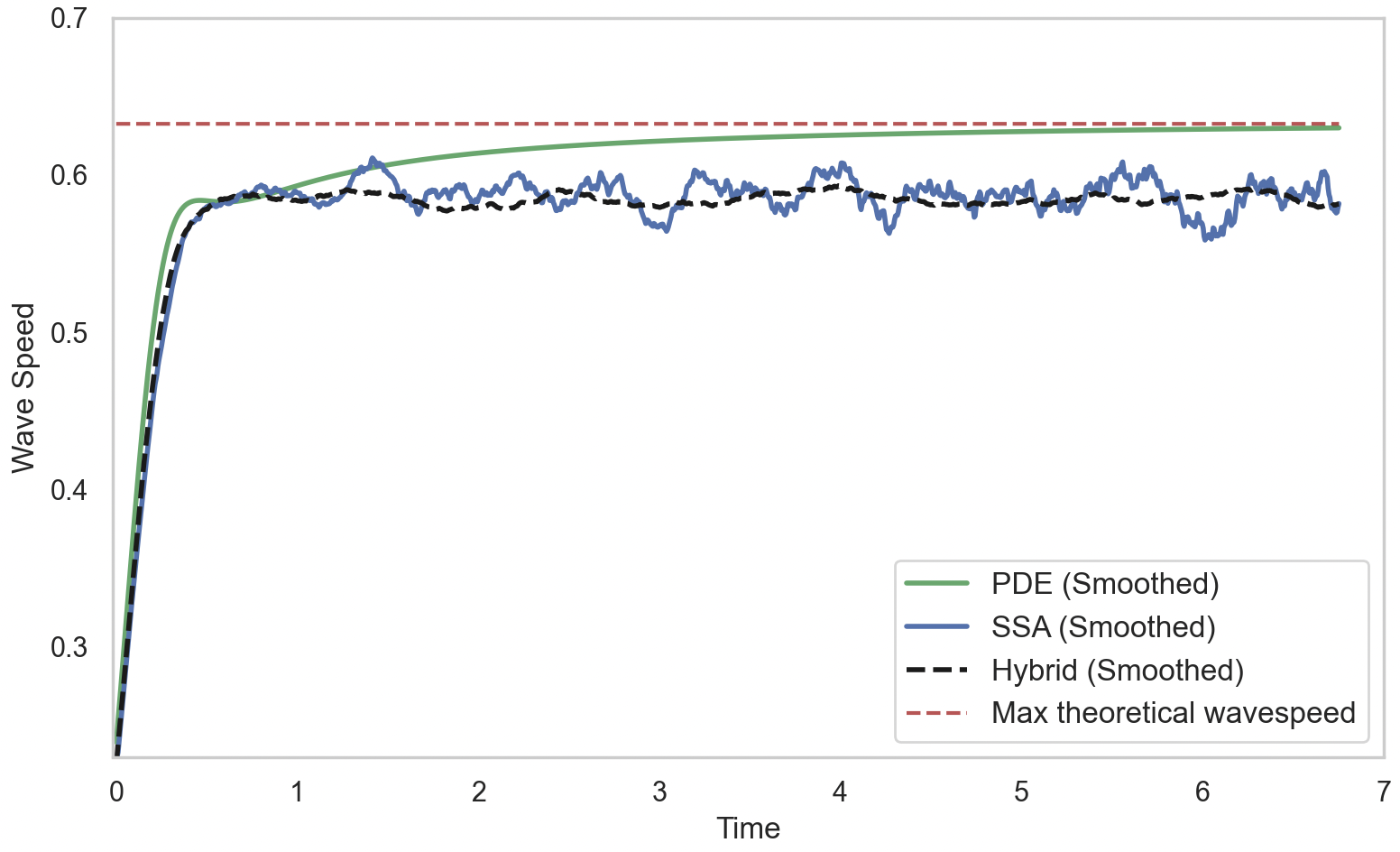}
        \caption{}
        \label{fig:FKPP_wavespeed}
    \end{subfigure}
    \hfill
    \begin{subfigure}[t]{0.48\textwidth}
        \centering
        \includegraphics[width=\linewidth]{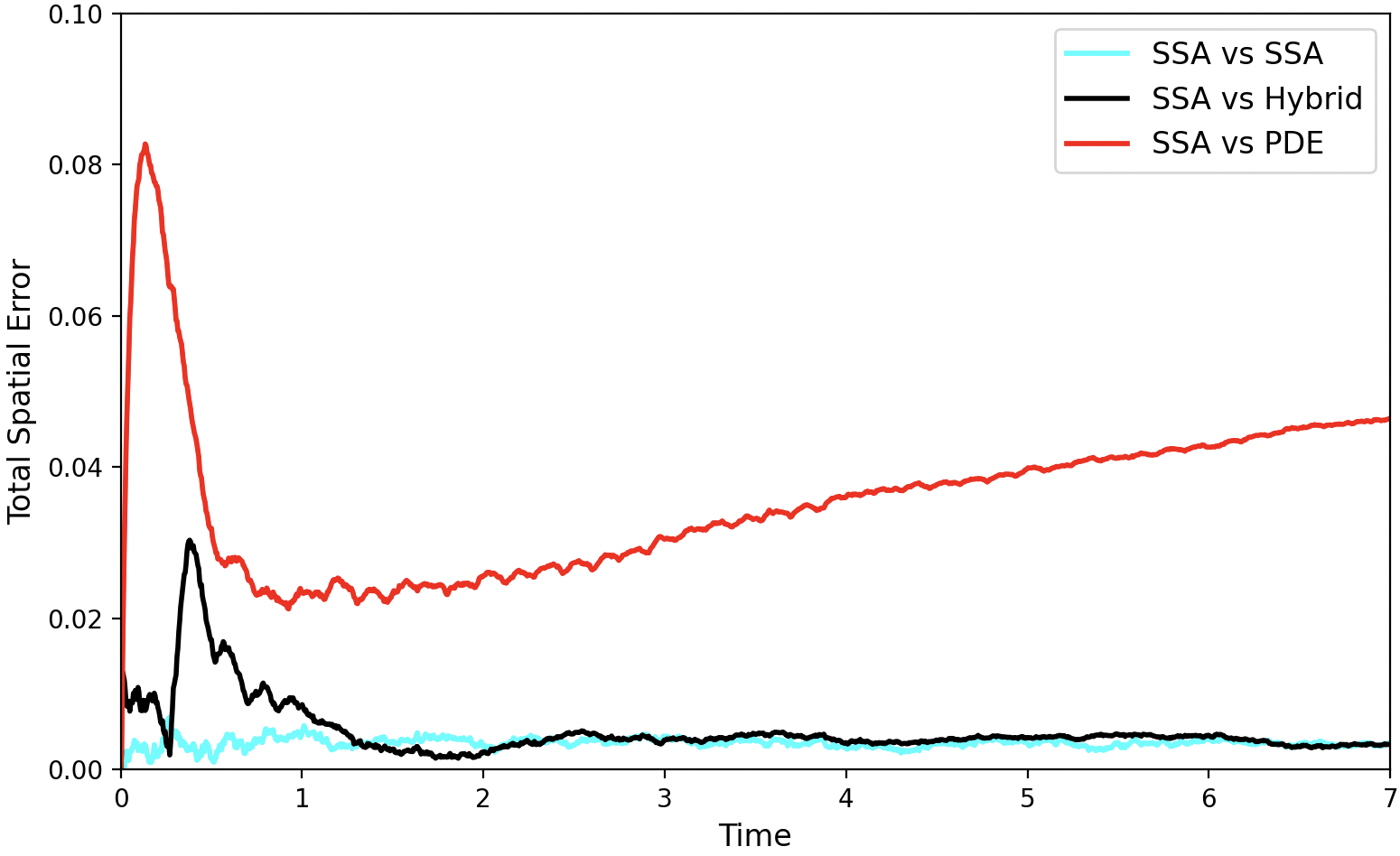}
        \caption{}
        \label{fig:fkpp_error}
    \end{subfigure}
    \caption{Comparison of (a) wave speed and (b) histogram distance errors errors across profiles averaged over 1000 repeats (stochastic models), where the SRCM is simulated using the adaptive production method. Parameters are as in figure \ref{fig:FKPP}.}
    \label{fig:fkpp_comparison}
\end{figure}

We observe in figure \ref{fig:FKPP} that there is now closer agreement between the stochastic and hybrid profiles for the adapted SRCM. Correspondingly, the smoothed wave speed of the SRCM now aligns much more closely with that of the canonical SSA simulation (see figure \ref{fig:fkpp_comparison} (\subref{fig:fig1_FKPP})).  This is to be expected, as the stochastic regime now dominates the dynamics at the leading edge of the wave - precisely where stochastic effects are important - allowing the wavefront dynamics to be governed primarily by the discrete stochastic model. As a result, the hybrid model captures the correct wave propagation speed of the canonical stochastic model. 

We observe that the error in the SRCM peaks early in the simulation and then settles to a level that closely matches the self-error of the SSA (see figure \ref{fig:fkpp_comparison} (\subref{fig:fkpp_error})). This initial spike is likely due to discontinuities in the concentration profiles during the early stages of wave formation, when particle numbers are low. 
The sharp transitions in the hybrid conversion mechanism in regions with minimal mass likely contribute to the transient peak in error.

\subsubsection{Efficiency}

We analyse the computational efficiency of modelling the FKPP system using both the canonical stochastic model and the SRCM (with adaptive production switching to ensure accuracy). Specifically, we compare the simulation run-times of these two methods when increasing the production rate, $\lambda_1$, from $1$ to $10$ (and consequently increasing the steady state value behind the wavefront) while keeping all other parameters the same.

The stable steady-state concentration of the reaction kinetics is given by the ratio of the production rate to the degradation rate. As the production rate increases, the total mass of the system increases proportionally. In the case of the pure SSA, this leads to a continually growing number of particles to simulate. Since the time step $\tau$ between reactions is inversely proportional to the sum of all reaction propensities, and this sum increases with the total number of particles, the SSA experiences an increase in time taken per repeat. In practice, this means that simulations become increasingly slow and eventually become computationally infeasible at high concentrations.

In contrast, the SRCM mitigates this issue through adaptive switching between stochastic and PDE representations. We fix the conversion threshold at 200 particles per unit volume. Once the local concentration exceeds this value, the system switches to a PDE representation in that region. As a result, even as the total mass increases, the SSA is only used to simulate a small proportion of the total mass, keeping the computational cost manageable.

\begin{figure}[H]
    \centering
    \includegraphics[width=0.9\linewidth]{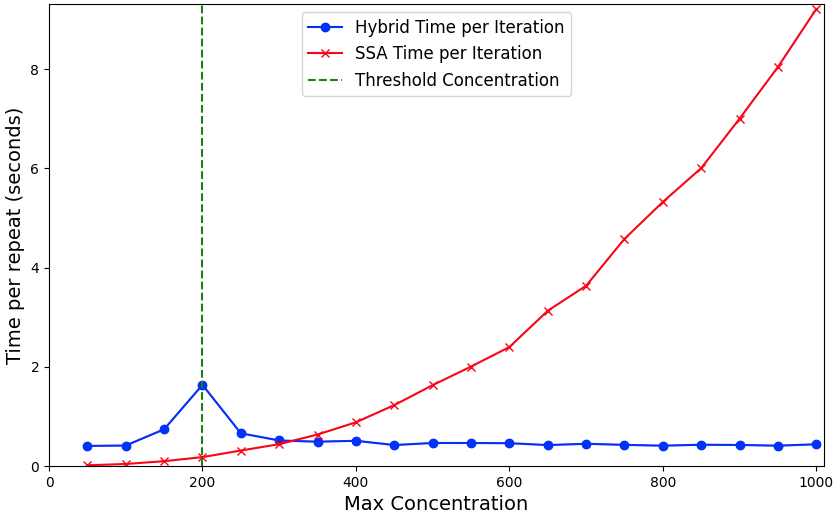}
    \caption{Average compute time per repeat for the FKPP system using the SRCM with adaptive production. Results are averaged over 10 repeats. The red line shows the runtime for the canonical SSA, while the blue line shows the runtime for the hybrid SRCM. Parameters are the same as figure \ref{fig:FKPP} with the exception of $\lambda_1$ which we vary.}
    \label{fig:efficiency}
\end{figure}

The dependence of the time taken per repeat on the production rate, $\lambda_1$, is illustrated in figure \ref{fig:efficiency}. Notably, for concentrations below the threshold, the time per repeat for the SRCM is larger than the canonical stochastic model as we pay all the costs of the stochastic simulation whilst also incurring the computational overheads of the SRCM. The SRCM peaks in time per iteration when the steady-state concentration is close to the switching threshold (200 particles per unit volume). Around this point, the concentration behind the wavefront fluctuates around the threshold, leading to many costly back-and-forth conversion events. This represents the least efficient instance for the SRCM and indicates that the conversion threshold should be chosen to be substantially below any expected steady-state concentration in order to avoid computational inefficiency.

As the production rate increases further and the steady-state concentration moves well beyond the threshold, the majority of the mass is represented by the PDE regime. The proportion of time spent simulating discrete stochastic mass becomes relatively smaller, improving the computational efficiency of the SRCM, in stark contrast to the ever-slowing canonical SSA. This highlights the practical limitations of the SSA for large-mass systems and the scalability advantages of the hybrid SRCM approach.


\section{Discussion}

In this work, we have described and illustrated the spatial regime conversion method, an adaptive hybrid modelling approach for reaction-diffusion systems. By dynamically allocating mass to either stochastic or deterministic regimes, based on local densities, the SRCM seeks to balance the accuracy of discrete stochastic simulations with the efficiency of continuous PDE models. Unlike traditional hybrid methods, requiring predefined or fixed interfaces \citep{Yates_Flegg_2015, Harrison_Yates_2016, Moro_2004}, the SRCM introduces local conversion reactions that facilitate the adaptive switching of regimes in response to evolving system states. Our simulation results demonstrate that this framework can substantially reduce computational cost while retaining key stochastic features, particularly in systems where mass varies substantially across time and space. However, while the SRCM performs well across a range of test cases, there are some subtleties to bear in mind.

In the pure diffusion test case (see section \ref{Section:Test_case_1}), the SRCM initially exhibited elevated error compared to the fully stochastic simulation. This was most pronounced during the early stages, where frequent conversion events introduced minor discrepancies in the SRCM concentration profile. While the overall error quickly converged to the same magnitude as the stochastic self-error, this early spike indicates a sensitivity to conversion timing. Smoother transitions or more gradual blending between regimes may mitigate this.

The morphogen gradient formation test problem (see section \ref{Section:Morphogen}) demonstrated the SRCM's ability to faithfully represent a spatial profile which crosses the threshold concentration. The hybrid model closely matched both the pure SSA and PDE outcomes. However, the hybrid error here remained marginally higher than the stochastic self-error across the simulation. This artifact may be a result of the difficulty of replicating the flux boundary condition perfectly in the SRCM, leading to a small but persistent discrepancy. 

In section \ref{Section:FKPP} we tested the SRCM on the FKPP travelling wave system, a classical benchmark for which front propagation is critically sensitive to stochastic effects. While the default SRCM captured wave formation and overall dynamics, its profile gradually diverged from the fully stochastic simulation. 
This discrepancy is fundamentally due to the fact that the wave speed is determined at the front, where stochastic effects should dominate; ideally, the entire wavefront region should be modelled using the discrete stochastic regime. However, in the originally specified implementation of the SRCM, the PDE governs a small but significant amount of the production within the front region. Even though this mass is quickly converted to discrete mass, this PDE-like behaviour at the very front overrides the stochastic dynamics that are critical to accurate wave propagation. As a result, the dynamics at the front are dictated by the PDE rather than the stochastic simulation, leading to an artificially accelerated wave. 

To correct this, we introduced an adaptive production mechanism, for which production occurs stochastically when the total local mass in a compartment is below a threshold and deterministically otherwise. This approach significantly improved the agreement between the SRCM wave speed and that of the canonical SSA benchmark, reducing the hybrid error. It makes sense to make this adaptive production mechanism the default for the SRCM. There is little point, when the total concentration is below the threshold, in producing mass in the PDE regime only for it to be rapidly converted into discrete mass. Indeed, as well as improving accuracy, this adaptation may also improve computational efficiency.

Future work will explore alternative frameworks for the threshold mechanism in the hybrid model. One potential avenue for improvement along this line is to introduce two separate thresholds at different concentration levels, rather than relying on a single threshold for both conversion reaction types. The current approach, which uses a single threshold, can lead to inefficiency due to conflicting conversions when the concentration profile fluctuates around to this conversion threshold, as illustrated in the efficiency plot (see figure \ref{fig:efficiency}). By contrast, with a two-threshold system, mass would convert from discrete to continuous when the total mass was above the higher threshold and from continuous to discrete when the total mass was below the lower threshold. This would create an intermediate concentration band where no conversion occurs, avoiding contiguity between the two conversion modes and reducing the risk of fluctuation-induced back-and-forth conversion.

A further extension worth investigating is replacing the sharp indicator function used to control conversions with a smoother sigmoidal function centred around a threshold. This would allow the conversion rate to vary continuously with concentration, making transitions more gradual.

The SRCM offers a versatile and extensible framework for modelling reaction-diffusion systems. It is particularly valuable when modelling multi-scale systems involving both low and high concentration regions.
This approach offers a promising foundation for modelling a wide range of systems for which switching behaviour, spatio-temporally varying concentration or front propagation play a critical role, including wound healing 
\citep{callaghan_stochastic_2006,Habbal_Barelli_Malandain_2014}, cell proliferation \citep{Vittadello_McCue_Gunasingh_Haass_Simpson_2018}, and biological pattern formation \citep{Klausmeier_1999, Sherratt_2013}. 

The potential to refine such schemes with smoother thresholds, to extend them to higher dimensions, and to employ them for real-world applications highlights an exciting avenue for future development. In general, this framework opens the door to handling a diverse range of multiscale systems robustly while significantly reducing computational costs.

\section*{Acknowledgments}

CGC is funded by the EPSRC Centre for Doctoral Training in
Statistical Applied Mathematics at Bath (SAMBa), under the project EP/S022945/1. 

The authors would like to acknowledge the initial helpful discussions with Dr. Martin Robinson, University of Oxford.

\section{Data availability}

The data presented in this study are openly available at \url{https://github.com/Cgyc20/SRCM_KPP}.

\bibliographystyle{unsrtnat}
\bibliography{references} 

\end{document}